\documentclass[prl,twocolumn,nobibnotes,a4paper]{revtex4-2}

\usepackage[pdftex]{graphicx}
\usepackage[pdftex]{epsfig}
\usepackage{amsmath}
\usepackage{amssymb}
\usepackage{amsfonts}
\usepackage{color}
\usepackage{wrapfig}
\usepackage{eucal}
\usepackage{hhline}
\usepackage{threeparttable}
\usepackage{supertabular}
\usepackage{multirow}
\usepackage{tabularx}
\usepackage{siunitx}
\usepackage{upgreek}
\usepackage{float}
\usepackage{ulem}
\usepackage{soul,xcolor}

\newcommand\numberthis{\addtocounter{equation}{1}\tag{\theequation}}

\usepackage[export]{adjustbox}

\usepackage{soul}

\usepackage[version=4]{mhchem} 
\usepackage{array} 
\usepackage[colorlinks=true, allcolors=blue]{hyperref} 

\newcommand{\und}[1]{_\textrm{#1}}

\begin{document}

\setstcolor{red}

\title{Non-classical mechanical states guided in a phononic waveguide}\thanks{This work was published in \href{https://doi.org/10.1038/s41567-022-01612-0}{Nature Phys.\ \textbf{18}, 789--793 (2022).}}

\author{Amirparsa Zivari, Robert Stockill, Niccol\`{o} Fiaschi, and Simon Gr\"oblacher}
\email{s.groeblacher@tudelft.nl}
\affiliation{Kavli Institute of Nanoscience, Department of Quantum Nanoscience, Delft University of Technology, 2628CJ Delft, The Netherlands}


\begin{abstract}
The ability to create, manipulate and detect non-classical states of light has been key for many recent achievements in quantum physics and for developing quantum technologies. Achieving the same level of control over phonons, the quanta of vibrations, could have a similar impact, in particular on the fields of quantum sensing and quantum information processing. Here we present a crucial step towards this level of control and realize a single-mode waveguide for individual phonons in a suspended silicon micro-structure. We use a cavity-waveguide architecture, where the cavity is used as a source and detector for the mechanical excitations, while the waveguide has a free standing end in order to reflect the phonons. This enables us to observe multiple round-trips of the phonons between the source and the reflector. The long mechanical lifetime of almost $\SI{100}{\micro s}$ demonstrates the possibility of nearly lossless transmission of single phonons over, in principle, tens of centimeters. Our experiment demonstrates full on-chip control over traveling single phonons strongly confined in the directions transverse to the propagation axis, potentially enabling a time-encoded multimode quantum memory at telecom wavelength and advanced quantum acoustics experiments.
\end{abstract}

\maketitle

Creating and detecting quantum states of mechanical motion open up new possibilities for quantum information processing, quantum sensing and probing the foundations of quantum physics~\cite{Aspelmeyer2014}. In particular within the field of quantum optomechanics many remarkable milestones have been reached over the past years:\ from showing the ability to realize the quantum ground state of a mechanical oscillator and single phonon control~\cite{OConnell2010}, unambiguous demonstration of the quantum nature of phonons through the creation of entangled states~\cite{Palomaki2013,Riedinger2018,Ockeloen-Korppi2018} and a Bell test~\cite{Marinkovic2018}, to realizing a long coherence-time quantum memory~\cite{Wallucks2020}. Furthermore, the field has shown the potential to enable crucial applications in connecting quantum computers, and transfer information between them~\cite{Andrews2014,Vainsencher2016,Forsch2020,Mirhosseini2020}. While these applications usually rely on highly confined phononic states (with a typical mode volume on the order of the wavelength), the use of traveling phonons promises the ability to create on-chip architectures for classical and quantum information~\cite{Habraken2012}, with the potential to add completely new capabilities compared to their optical counterparts. The exciting prospects in this new field of quantum acoustics are enabled by the orders of magnitude slower propagation speed compared to photons, the inherently low loss, their extremely low energy and the small mode volume compared to GHz-frequency photons. These features make phonons ideally suited for direct manipulation on a chip with wavelength sized components, while the ability to realize significant time-delays in a short distance makes this type of system an ideal platform for on-chip operations~\cite{Delsing2019}. Furthermore, phonons have also demonstrated their unique capability to efficiently couple and even mediate the interaction between various quantum systems~\cite{Schuetz2015}, such as superconducting qubits~\cite{Bienfait2019}, defect centers in solids~\cite{Golter2016}, and quantum dots~\cite{Hermelin2011,McNeil2011}. Using mechanical excitations as low-loss carriers of quantum information will allow for the construction of two-dimensional architectures and large-scale phononic quantum networks~\cite{Kuzyk2018}.

While the creation of non-classical mechanical states have been demonstrated in multiple physical systems~\cite{Chu2020}, only a very limited number of experiments have been able to realize propagating modes in the quantum regime, all based on surface acoustic waves~\cite{Gustafsson2012,Gustafsson2014,Bienfait2019}. This approach comes with its own limitations and challenges, such as relatively short lifetimes, losses due to beam steering and diffraction, typically only bi-directional emission and no full confinement of the mode except in resonators. In the classical domain on the other hand, several proof-of-concept experiments have realized the creation, transport and detection of mechanical states at cryogenic temperatures~\cite{Patel2018} over millimeter ranges, as well as at room temperature and atmospheric pressure~\cite{Fang2016}. The possibility of guiding a mechanical quantum state in a waveguide that confines the excitation in all directions transverse to its propagation, similar to optical fibers and waveguides, remains an open challenge.

\begin{figure*}[ht!]
	\includegraphics[width = \linewidth]{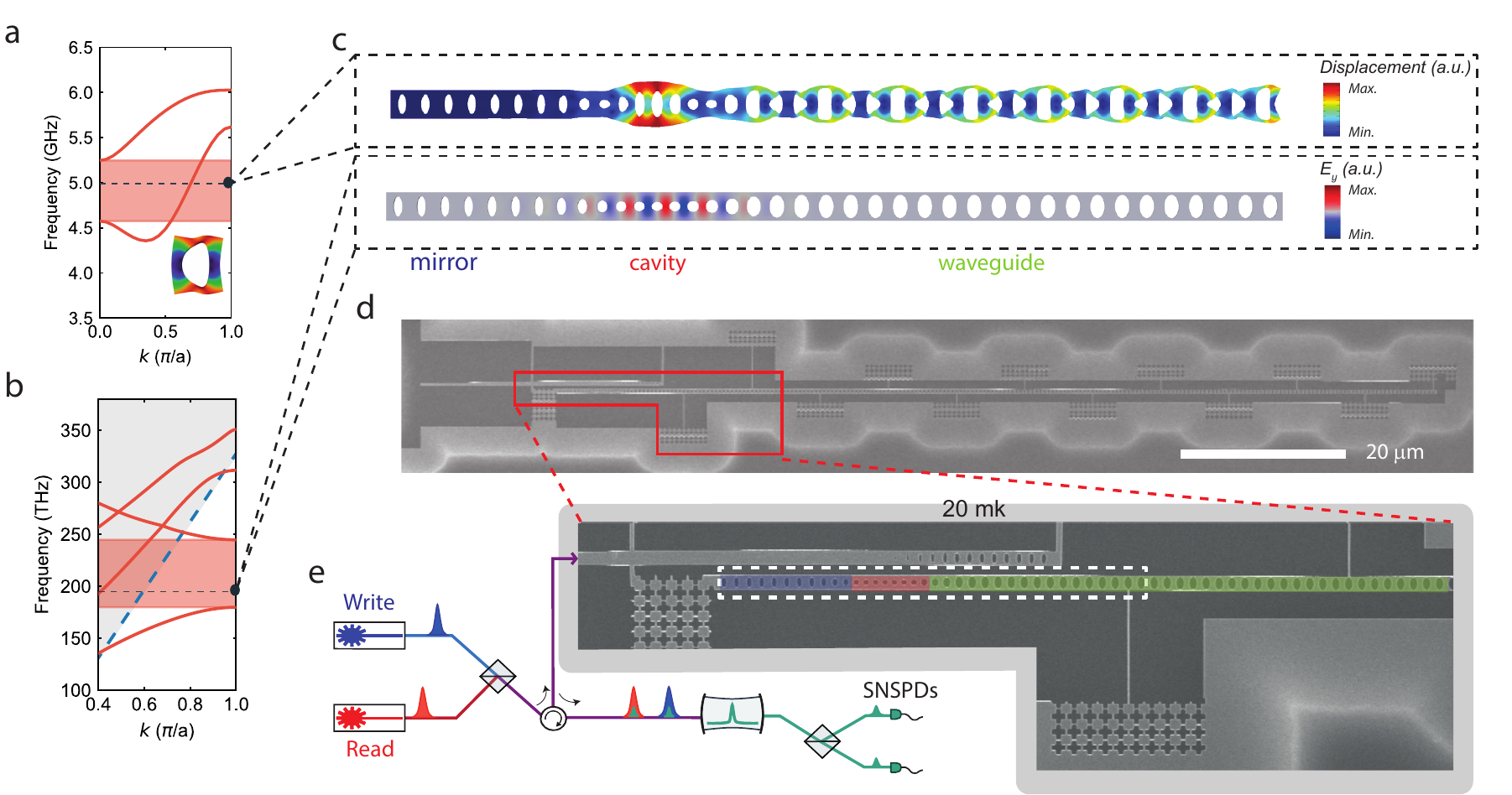}
	\caption{a) Band diagram for the modes with symmetric displacement field with respect to the propagation direction along the waveguide. These modes are expected to couple efficiently to the resonant optomechanical cavity mode. The highlighted region (red) shows a single mode waveguide for the symmetric breathing mode with linear dispersion. The dashed line represent the frequency of the modes of the optomechanical structure. Inset:\ mode shape of the unit cell from this simulation for the band of interested. b) Band diagram for the optical TE mode of the phononic waveguide, exhibiting a band gap at telecom wavelengths, allowing for a confined optical mode inside the optomechanical resonator. The dashed blue line is the light cone, with the non guided modes in the gray shaded area. c) Mechanical (top) and optical (bottom) eigenfrequency simulation of the full cavity and waveguide structure. Clearly visible are the resonant breathing mode of the optomechanical crystal (left) and the waveguide mode (right). As the mechanical mode leaks through the waveguide, the optical mode stays confined inside the optomechanical resonator. d) A scanning electron microscope image of the device used in the experiment, showing the full device with mirror, cavity and the \SI{92}{\mu m} long waveguide, as well as the optical coupling waveguide (top left). e) Schematic of the setup together with a zoomed-in section of (d) (indicated by the red box). The blue, red and green shaded regions show mirror, cavity and waveguide, respectively. The white dashed rectangle is the area of the simulation in (c). See the text and SI~\ref{SI} for a more details.}
	\label{Fig:1}
\end{figure*}

Here we demonstrate a single-mode phononic waveguide directly coupled to an on-chip source and detector for non-classical mechanical states. We verify the non-classicality of the launched mechanical states by measuring their quantum correlations with an optical read-out field. In particular, we use the optomechanical interaction to herald the creation of a single phonon, which then leaks into the phononic waveguide. Since the waveguide has a free standing end that acts as a mirror for the phonons, the excitations bounce, i.e.\ reflect, back and forth with a certain characteristic time that is determined by the group velocity and the length of the waveguide. Moreover, we observe non-classical correlations between time-bin encoded phonons~\cite{Farrera2018}, by creating and detecting a phonon in either an early or late time window. The long mechanical lifetime of the device will also allow the creation of an on-chip network for quantum acoustic experiments. With the on-chip source, detector and waveguide presented in this work, only a phononic beam splitter and phase modulator need to be developed in order to obtain full coherent control over phonons on a chip.

We design our phononic crystal waveguide in thin-film silicon, which is single-mode for the symmetric breathing mode of the structure, in the frequency range of interest (at around $\SI{5}{GHz}$ with a single mode range of $\SI{750}{MHz}$) and has an approximately linear dispersion, in order to maintain the spatial mode shape of the traveling phonons. This waveguide is connected to an optomechanical resonator acting as the single-phonon source and detector. For the waveguide design, only the symmetric breathing mode is considered, in order to enable a good mode overlap with the mode of the optomechanical cavity, as these resonant modes have large optomechanical coupling, and can easily be created and detected optically. At the same time, in order to realize a high-finesse optical cavity, we design our phononic waveguide to act as a mirror for photons, therefore confining the optical field in the optomechanical resonator.

The details of our design are shown in Fig.~\ref{Fig:1}a and \ref{Fig:1}b, where we plot the band structure with the right mechanical symmetry, as well as the transverse electric (TE) polarized optical mode. An eigenvalue simulation of the full structure, cavity and waveguide, is shown in Fig.~\ref{Fig:1}c. The mechanical mode extends into the waveguide, while the optical mode is strongly confined to the cavity region with a similar mode volume to previous works~\cite{Chan2012,Hong2017,Riedinger2018,Qiu2020}. The different sizes of the holes in the structure create the mirror, defect and waveguide.
The hole dimensions and periodicity in the waveguide part are adjusted to tune the group velocity. We design the waveguide to have a small group velocity while still having a linear band inside the frequency range of interest (see SI for more details).
From the simulated group velocity, the time duration of the mechanical packet (set by the optical pulse length of $\SI{40}{ns}$), and the time of the mechanical excitation to leak from the cavity to the waveguide, we determine a minimum length of about $\SI{40}{\micro m}$ for the waveguide for the excitation to completely leave the cavity before it comes back again, which is why we choose a length of $\SI{92}{\micro m}$.
In order to support the long waveguide after suspension, we use narrow (\SI{50}{nm} wide) tethers to connect it to the surrounding silicon. Moreover, to prevent any mechanical dissipation through the tethers, they are directly connected to a phononic shield, as can be seen in the zoomed-in image of the device in Fig.~\ref{Fig:1}e. The phononic shield features a bandgap from \SI{4}{GHz} to \SI{6}{GHz}, and by increasing the number of periods in the shields we can increase the mechanical lifetime (see SI). The same phononic shields are used at the left end of the device, to further increase the mechanical lifetime.

A picture of the device and sketch of the experimental setup can be found in Figure~\ref{Fig:1}d-e. To excite and detect non-classical phonons we use laser pulses detuned from the optical resonance to address the optomechanical Stokes and anti-Stokes sidebands in order to create (write) the mechanical excitation and to map it onto the optical mode (read), respectively~\cite{Riedinger2016}. After being combined on a 50:50 beam splitter (BS), the light is routed via an optical circulator to the device. The reflected light from the device is then filtered using free space Fabry-P\'erot cavities to block the pump laser pulses and, after another BS, is sent to the superconducting nanowire single photon detectors (SNSPDs). The device itself is cooled to $\SI{20}{mK}$ in order to initialize the mechanical mode of interest deep in the ground state.

\begin{figure}
	\includegraphics[width = 1 \linewidth,left]{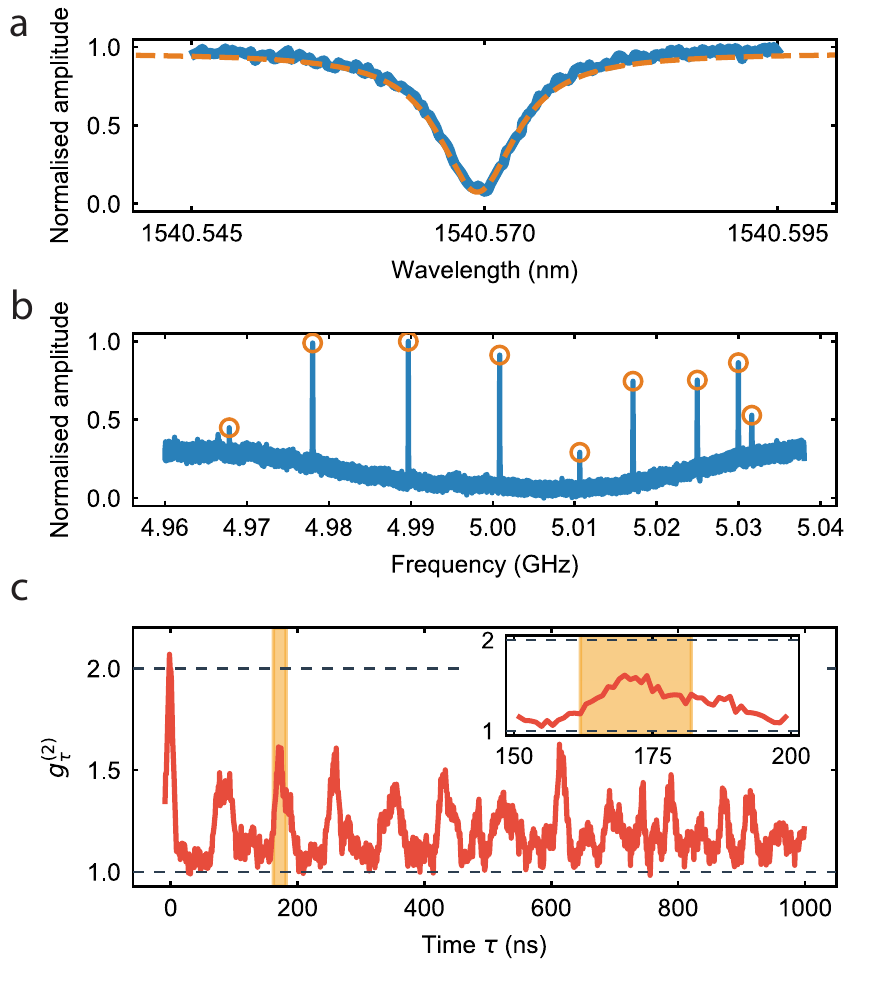}
	\caption{(a) Characterization of the optical resonance of the device in reflection. (b) Mechanical spectrum measured using the OMIT technique at $\SI{20}{mK}$. The series of peaks is given by the hybridization of the cavity mode and the modes of the waveguide (approximately equally spaced). (c) $g^{(2)}_{\uptau}$ of a waveguide-coupled thermal state for different delays between two detection events ($\uptau$). Note the series of peaks indicating the traveling back and forth of the phonons in the waveguide. The reduced maxima for these peaks is attributed to the non constant FSR. The area highlighted is the chosen round trip peak for the pulsed experiment with single phonon states. Inset:\ zoom-in around the highlighted area.}
	\label{Fig:2}
\end{figure}

For the initial characterization of the device, we use a tunable continuous-wave (CW) laser to determine the optical resonance in reflection. As shown in Fig.~\ref{Fig:2}a, the fundamental optical resonance has a central wavelength around \SI{1541}{nm} and a linewidth of $\kappa\und{t} = \SI{1021}{MHz}$ (with extrinsic and intrinsic loss rates of $\kappa\und{e} = \SI{364}{MHz}$ and $\kappa\und{i} = \SI{656}{MHz}$). We measure the mechanical spectrum using the optomechanical induced transparency (OMIT) technique~\cite{Weis2010} (see SI for details). The resulting renormalized amplitude of the reflected probe field ($|S_{21}|$) is plotted in Fig~\ref{Fig:2}b, where a series of (almost) equally spaced peaks shows the hybridization of the single mode of the cavity with the series of modes of the free-ended waveguide. We choose the first prominent mechanical resonance (around \SI{4.98}{GHz}) as the frequency to which we detune the laser with respect to the optical resonance wavelength for addressing the Stokes and anti-Stokes interaction. We further measure the equivalent single photon optomechanical coupling rate from the Stokes scattering probability using a short optical pulse of full width at half maximum (FWHM)~$\approx\SI{40}{ns}$, obtaining a collective $g_{0}/2\pi \approx \SI{460}{kHz}$ (for a detailed explanation see the SI). This is the joint coupling rate of all mechanical eigenmodes within the detection filter bandwidth.

In order to determine the time dynamics of the phononic wave packet, we measure the second order correlation function $g^{(2)}_{\uptau}$ of the light scattered from the cavity with a CW read-out tone detuned to the anti-Stokes sideband. Due to non-negligible optical absorption in silicon, the continuous laser creates thermal mechanical population in the device~\cite{Meenehan2014}.
In this experiment, the continuous red-detuned laser field excites thermal phonons in a broad frequency range. These phonons are read out by the same red-detuned field, which allows us to only measure the anti-Stokes scattered photons on resonance with the optical cavity. The phononic state is therefore mapped onto the photonic state, and the photons are finally detected using our SNSPDs. As a result the photon statistics of the optical field corresponds to the phonon statistics of the thermal mechanical mode.
We obtain the $g^{(2)}_{\uptau}$ between emitted photons from the device, by measuring two photon coincidences on two different SNSPDs and normalizing them to the single photon counts of the SNSPDs. Owing to the optomechanical interaction, this is equivalent to measuring the $g^{(2)}_{\uptau}$ of the mechanical thermal states, for different delays between the SNSPDs clicks $\uptau$. The results are plotted in Fig~\ref{Fig:2}c, showing an (almost) equally spaced series of peaks. As expected for a thermal state, at $\uptau=0$, we observe $g^{(2)}_{\uptau}=2$, which is then modulated as the state leaks into the waveguide. We attribute the reduced maxima for the round trip peaks ($g^{(2)}_{\uptau}\approx 1.5$ instead of 2) to coupling to undesirable asymmetric mechanical modes (which have relatively low optomechanical coupling rate, and thus cannot be detected optically), as well as the non-constant FSR between mechanical modes of the device, which could be caused by the dispersion of the phononic waveguide. The exact effect will require a more detailed theoretical and experimental analysis in the future.

We use three 150-MHz broad filter cavities in series to filter out the strong optical driving pumps, which also filter the Stokes and anti-Stokes scattered photons within a frequency range of $\SI{80}{MHz}$ around the setpoint (\SI{4.98}{GHz}). In this way, any signal from the mechanical modes greater than \SI{5.02}{GHz} is strongly suppressed and hence, only the part of the spectrum with the evenly spaced mechanical modes will contribute considerably to the correlation. These modes build a frequency comb with a free spectral range (FSR) of around \SI{11}{MHz}, which corresponds to a rephasing time of $1/\textrm{FSR} = \SI{91}{ns}$. This is consistent with the round trip time that can be inferred from the measurement, which is around \SI{85}{ns}.

In order to verify that we can guide a non-classical mechanical state, we employ a scheme in which we herald the creation of a quantum excitation in the optomechanical cavity, which we confirm by swapping out the mechanical excitation to an optical photon after some time and correlating the photon statistics from the two processes~\cite{Riedinger2016}. We obtain these correlations by measuring the coincidences between the events on the SNPSDs. We realize this scheme by first addressing the Stokes process with a $\SI{119}{fJ}$ blue-detuned $\SI{40}{ns}$ laser pulse creating a two-mode squeezed opto-mechanical state with scattering probability of $p\und{s,write} \approx 1.4 \%$. Similarly, to readout the mechanical state from the optomechanical cavity, a red-detuned laser pulse with the same energy, duration and scattering probability of $p\und{s,read} \approx 1.4 \%$, is sent to the device addressing the anti-Stokes process. These low values are chosen to avoid excess heating of the optomechanical device from the remaining optical absorption in the silicon. Note however that it has been shown that these can be increased up to around $\SI{30}{\%}$~\cite{Hong2017}. These scattering probabilities set the thermal occupation of the mode of interest to $n\und{th}\approx\SI{0.27}{}$ (see SI).

\begin{figure}[t!]
	\includegraphics[width = 1\linewidth]{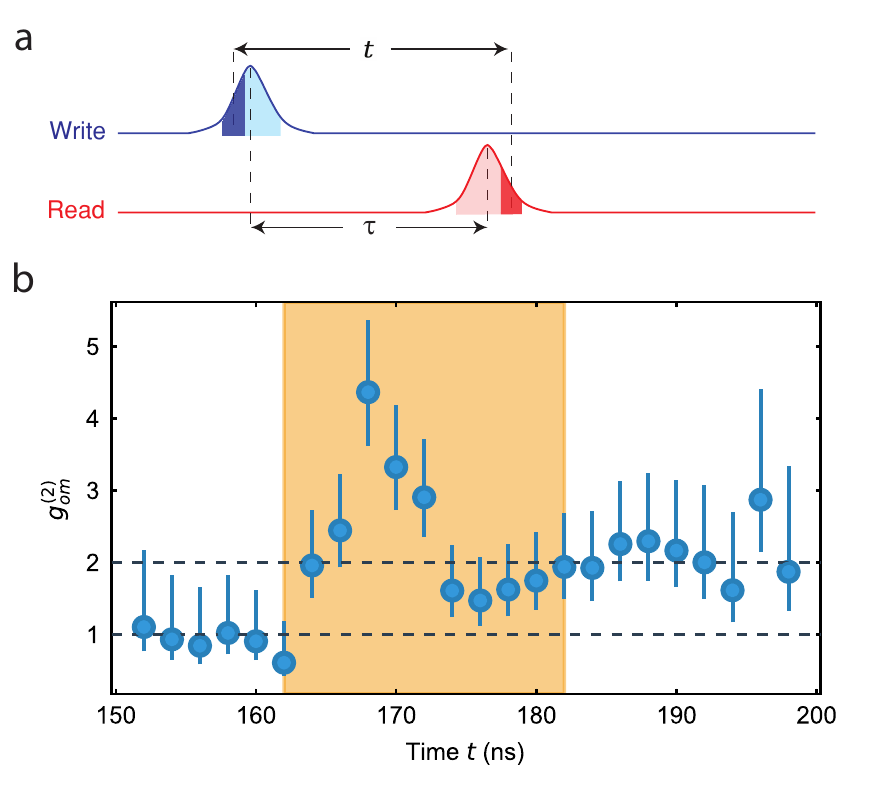}
	\caption{a) Pulse scheme used for the cross correlation measurement. We fix the time between the pulses to $\uptau \approx \SI{170}{ns}$, as calibrated from the measurement of Fig~\ref{Fig:2}c. In post-processing, we scan a narrow time window of \SI{6}{ns} with an adjustable delay of $t$ represented by the dark blue and red shaded areas over all of the acquired data (light shaded areas). This approach allows us to calculate the $g^{(2)}\und{om}$ of each point with an adjustable moving window. b) Cross correlation between the optical and the mechanical state for the pulsed experiment $g^{(2)}\und{om}(t)$. The correlation is higher than the classical threshold of 2, with a maxima of $g^{(2)}\und{om} = 4.4^{+1.0}_{-0.7}$, clearly demonstrating the non-classical character of the traveling phonons. For more details on the error calculation see the SI~\ref{SI}. The highlighted area in the figure has the same position as the one in Fig~\ref{Fig:2}c.}
	\label{Fig:3}
\end{figure}

The scheme of the pulses can be seen in Fig.~\ref{Fig:3}a and the delay between the red-detuned and blue-detuned pulses is set to approximately the second round trip time ($\uptau\approx\SI{170}{ns}$). The detection of a single photon in one of the detectors from the blue-detuned pulse heralds the mechanical state of the defect to a single-phonon Fock state~\cite{Hong2017}. This phonon leaks through the attached phononic waveguide after a short time $T\und{c}$ and travels back and forth between the defect part and the end of the waveguide. In Figure~\ref{Fig:2}c the highlighted area shows the chosen peak, that has a delay of $\uptau\approx\SI{170}{ns}$, which is much smaller than the measured lifetime $T_1\approx\SI{78}{\micro s}$ for this particular device (see the SI for more details). We choose to perform the measurement after two round trips of the phonons to avoid any overlap between the optical write and read pulses. Another reason for this is to overcome the SNSPDs' dead time of around $\SI{100}{ns}$ and hence be able to measure coincidences on the same SNSPD as well as from different SNSPDs. We would like to note that the expected cross-correlation between phonons and photons on multiple round trips is expected to be similar or slightly lower as a result of optical absorption and delayed heating~\cite{Riedinger2016,Hong2017}. From this measurement, we can also infer the coupling between the cavity and the waveguide from the width of the chosen peak, obtaining a decay time of $T\und{c}\approx\SI{10}{ns}$, corresponding to a coupling rate of around $2\pi\times\SI{16}{MHz}$.

In order to measure the coincidences required to determine the correlations, two 6-ns wide time windows, with a varying delay $t$ between them, are scanned through the whole area of pulses in a post processing step (see Fig.~\ref{Fig:3}a). Note that we also use the measurement shown in Fig.~\ref{Fig:2}c to calibrate these filtering windows. By summing all the coincidences with each click happening in the same trial ($\Delta n = 0$, for $n$ indicating each trial), we gather all the correlated coincidences for each delay. In order to obtain the average uncorrelated coincidences, we perform a similar post processing step, but finding coincidences in different trials ($\Delta n \neq 0$) where the clicks can be assumed to be uncorrelated. Averaging this value over different $\Delta n \neq 0$, gives the average number of uncorrelated coincidences. We use these two values to calculate $g^{(2)}\und{om}(t)$. The time window is chosen to be less than the coupling time between the cavity and the waveguide ($T\und{c}$) in order to select only the correlated photons that have indeed traveled in the waveguide (see SI for more details). The result is shown in Fig.~\ref{Fig:3}b. In order to gather more statistics, two separate measurements with identical thermal occupation have been used for the data shown here, one with the sequence of red and blue pulses repeated every $\SI{200}{\micro s}$, the other every $\SI{300}{\micro s}$. After merging all the coincidences, a maximum cross-correlation of $g^{(2)}\und{om} = 4.4^{+1.0}_{-0.7}$ is obtained from the reflected phonons at a time of $t=\SI{168}{ns}$, which is more than $3$ standard deviations above the classical threshold of $2$, unambiguously showing the non-classical behavior of the guided single phonon state \cite{Riedinger2016}. Furthermore, no non-classical correlations between the photons can be observed at times where the phonon is not spatially located inside the defect (i.e.\ outside a window of width $T\und{c}$ centered at $t \approx \uptau \approx\SI{170}{ns}$). We suspect the slightly increased correlations at longer times to be a result of the waveguide dispersion. This effect is also clearly visible in the envelope of the peak shape corresponding to the second round trip in the inset of Fig.~\ref{Fig:2}c, with both patterns closely resembling one another. 

\begin{figure}[h!]
	\includegraphics[width = 1\linewidth]{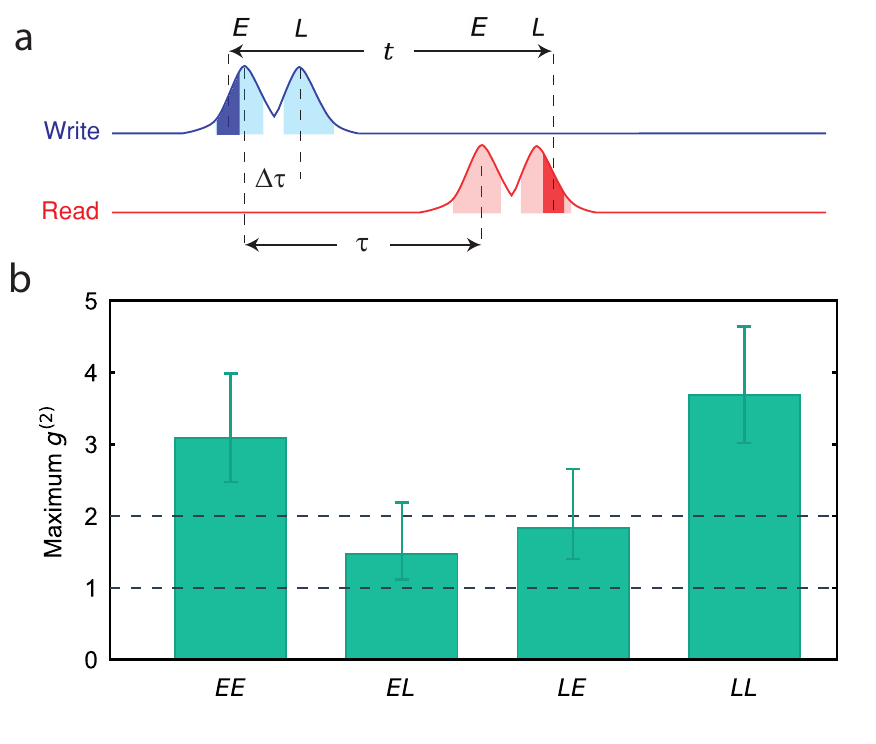}
	\caption{a) Pulse scheme for the generation of time-bin encoded phonon states with $\Delta\uptau=\SI{45}{ns}$ and $\tau = \SI{170}{ns}$. The area used to gather coincidences is depicted by the light blue (red) shaded area for the write (read) pulses, respectively. In the post-processing step a similar technique as before is used, in order to achieve high timing resolution. An example of $\SI{6}{ns}$ wide coincidence-filtering windows are sketched by dark blue (red) areas for write (read) pulses for the "early-late" combination. $E$ ($L$) stands for early (late). b) As expected, the maximum cross correlations for the various settings clearly shows non-classical correlations for the "early-early" and "late-late" combinations, whereas in the other two we only observe classically correlated phonons.}
	\label{Fig:4}
\end{figure}

We further explore the potential of creating a time-bin encoded phononic state by extending our scheme to using two optical excitation (write) and two detection pulses (read), effectively realizing a phononic FIFO quantum memory~\cite{Pang2020}. The identical blue-detuned optical pulses (FWHM $\SI{40}{ns}$) have a scattering probability of $p\und{s,write} \approx \SI{2.7}{\%}$ with $\Delta \tau = \SI{45}{ns}$ delay. The red-detuned detection pulses (FWHM $\SI{40}{ns}$) have a scattering probability of $p\und{s,read} \approx \SI{1.5}{\%}$ and with the same time delay. The first blue- and red-detuned pulses are spaced by $\tau = \SI{170}{ns}$ from each other and we repeat this sequence every $\SI{800}{\mu s}$. We then measure the maximum cross correlation in time between all four combinations of write and read pulses, using the same technique for delay-filtering used to extract the data shown in Fig.~\ref{Fig:3}.
We choose 40-ns-long non-overlapping time windows to separate "early" write and read pulses from "late" ones, as depicted in Fig.~\ref{Fig:4}a.
As shown in Fig.~\ref{Fig:4}b, we can clearly see strong non-classical correlation between the "early-early" and "late-late" combination of excitation and detection pulse, while observing only classical correlations (due to absorption-induced heating) between the other combinations of "early-late" and "late-early".

Our results clearly demonstrate the potential for creating, guiding, and detecting a non-classical mechanical state inside a phononic crystal, using optomechanical techniques. Thanks to the long mechanical lifetime (up to \SI{5.5}{ms} for similar devices on the same chip) and the full lateral confinement, this type of device paves the way towards on-chip quantum acoustic experiments. The current efficiency of the device is limited by residual optical absorption in the silicon, which can be reduced through improved fabrication~\cite{Hong2017} and surface passivation~\cite{Borselli2006}. While some of the motivation for these experiments stems from the similarity to quantum optics, phonons are crucially different, due to the 5 orders of magnitude smaller propagation speed and the ease of coupling them to other quantum systems. The realization of phononic beamsplitters and phase modulators will complete the toolbox required for full control over traveling single phonons and more complex quantum experiments on a chip.
Furthermore, the possibility of retrieving the state after several round-trips and with having time-bin encoded phononic states, together with the full engineerability of the bandstructure, will allow for the creation of time-bin encoded phononic qubits and an optomechanical multimode quantum memory working natively at telecom wavelengths. We also expect guided phonon modes to play a crucial role in the low-energy transmission of quantum information on a chip and for next generation filtering.

\begin{acknowledgments}
	We would like to thank Ewold Verhagen and Roel Burgwal for valuable discussions, as well as Moritz Forsch for experimental support. We further acknowledge assistance from the Kavli Nanolab Delft. This work is financially supported by the European Research Council (ERC CoG Q-ECHOS, 101001005), and by the Netherlands Organization for Scientific Research (NWO/OCW), as part of the Frontiers of Nanoscience program, as well as through Vidi (680-47-541/994) and Vrij Programma (680-92-18-04) grants. R.S.\ also acknowledges funding from the European Union under a Marie Sk\l{}odowska-Curie COFUND fellowship.
\end{acknowledgments}


%

\setcounter{figure}{0}
\renewcommand{\thefigure}{S\arabic{figure}}
\setcounter{equation}{0}
\renewcommand{\theequation}{S\arabic{equation}}

\clearpage

\section{Supplementary Information}
\label{SI}

\subsubsection{Fabrication}
The device is fabricated from a silicon-on-insulator chip with a device layer of \SI{250}{nm}. We first pattern the structure using electron-beam lithography into the resist (CSAR) and then transfer it to the silicon layer with a $\ce{SF6}$/$\ce{O2}$ reactive-ion etch. The structure is then cleaned with a piranha solution and suspended by undercutting the sacrificial oxide layer using hydroflouric acid (HF). A last cleaning step with piranha and diluted HF (1\%) is performed just before mounting the sample in the dilution refrigerator.

\subsubsection{Experimental setup}
\begin{figure}
    \includegraphics[width = 1\linewidth]{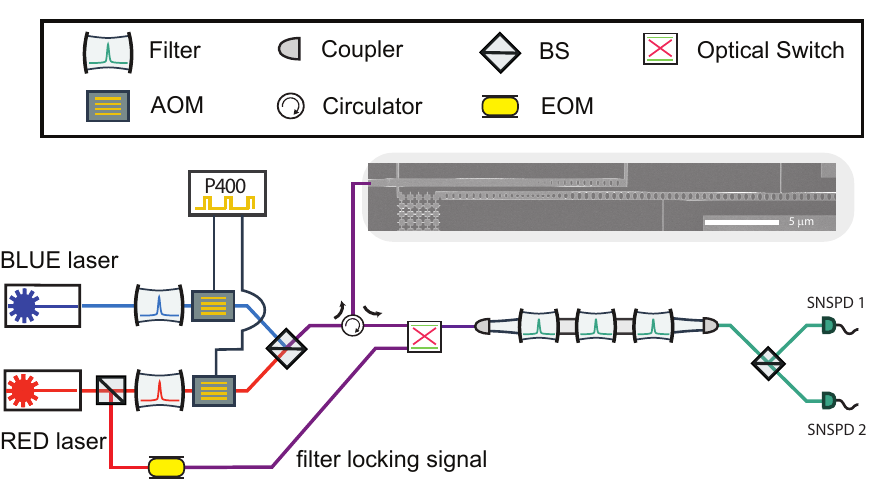}
    \caption{The experimental setup, with BS the beam splitter and SNSPD the superconducting nanowire single-photon detectors. See text for more details.}
    \label{Fig:S1}
\end{figure}

A complete overview of the experimental setup is shown in Fig.~\ref{Fig:S1}. The optical control pulses are created from two separate tunable CW external-cavity diode lasers. They are stabilized over time using a wavelength meter (not shown) and are filtered with fiber based filter cavities (\SI{50}{MHz} linewidth) to suppress the classical laser noise at GHz frequencies. We create the pulses using two \SI{110}{MHz} acousto-optic modulators (AOMs) gated via a pulse generator. The optical paths are combined through a fiber beamsplitter. The pulses are routed to the device in the dilution refrigerator via an optical circulator, and then to the device's optical waveguide using a lensed fiber. The optical signal coming from the device is filtered with three $\SI{150}{MHz}$ broad free-space Fabry-P\'erot cavities in series, that results into an equivalent series bandwidth of about $\SI{80}{MHz}$. The average suppression ratio is \SI{114}{dB} for the blue detuned pulse and \SI{119}{dB} for the red detuned pulse (the small difference is due to small misalignment and mode mismatch of the cavities). Every ${\sim}\SI{10}{s}$ the experiment is paused and continuous light is send to the filter cavities to lock them at the optical resonance of the device using optical switches. For this, the red-detuned laser is used after passing through a 50:50 beam splitter, and the detuning is compensated via an electro-optics modulator to match the optical resonance. It takes around $\SI{6}{s}$ to lock the cavities each time.

The effect of dark counts, mostly unwanted photons from stray light which couple into the fiber and leakage of pump photons through the filter chain, are negligibly small compared to the photon rate that is obtained via the Stokes and anti-Stokes scatterings. During the cross-correlation measurement shown in Fig.~\hyperref[Fig:3]{\ref{Fig:3}}, we get total photon rates of approximately $8\times 10^{-4}$ and $1.7\times 10^{-4}$ per trial from the Stokes and anti-Stokes scatterings, respectively, from the full area of optical pulses. At the same time, the average total dark count rate, which is constant in time, is only $1\times 10^{-5}$ per trial.

\subsubsection{Coherent excitation measurement}

\begin{figure}
    \includegraphics[width = 1\linewidth]{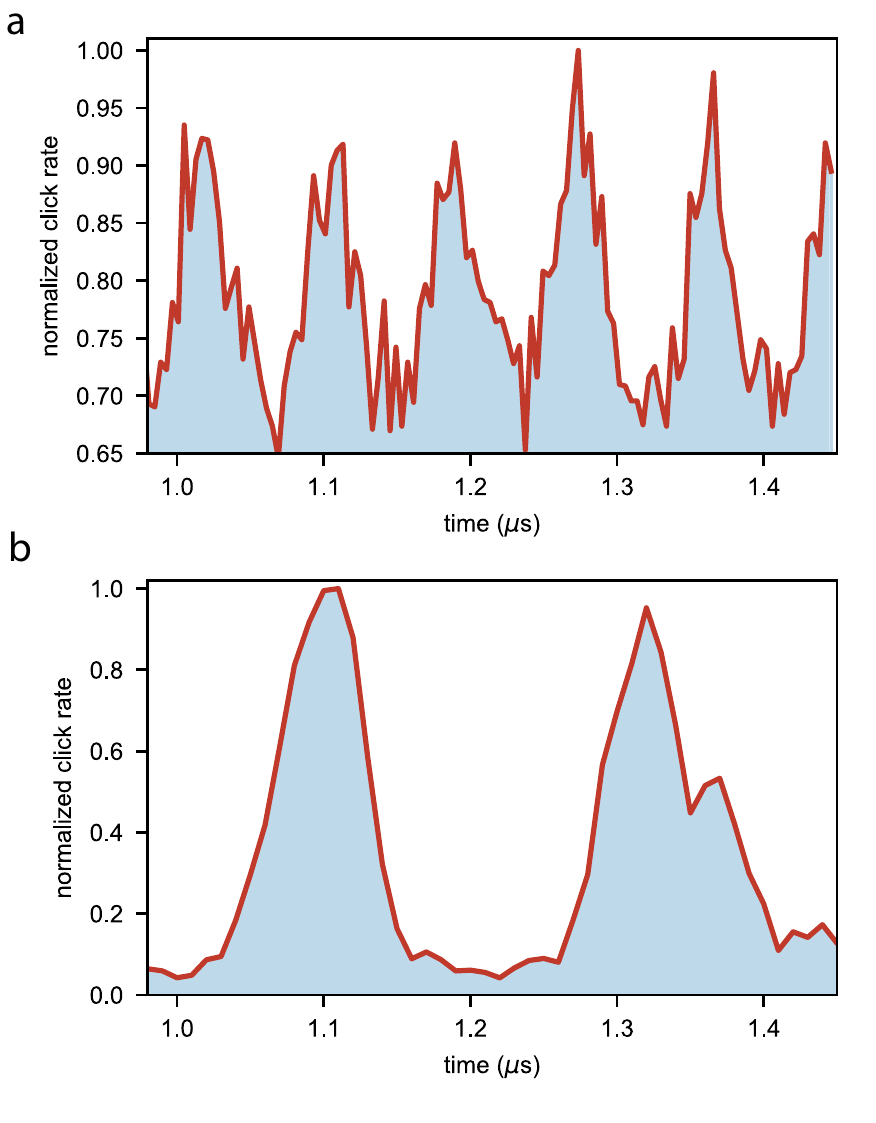}
    \caption{Coherent mechanical population of the cavity versus time for different waveguide lengths of a) \SI{92}{\mu m} and b) \SI{230}{\mu m}. The time difference between the two peaks in the bottom figure is around $\tau \approx \SI{210}{ns}$ compared to $\tau \approx \SI{85}{ns}$ for the shorter waveguide, which is fully consistent with the difference in waveguide length.}
    \label{Fig:coh_drive}
\end{figure}

Here we study the routing of a coherent mechanical wavepacket, as a verification to the round-trip time of the measurement of Fig.~\ref{Fig:2}c of the main text. We excite the mechanical wave packet with a \SI{40}{ns} long pulse of laser light, blue-detuned by \SI{7}{GHz} with respect to optical resonance of the cavity. Importantly, as the Stokes scattered photons are more than two linewidths away from the optical resonance, the Stokes process will be highly attenuated. We pass this laser pulse through an electro-optical modulator (EOM) which produces two side bands at around \SI{4.978}{GHz} (the frequency of the mechanical mode which we use to detune our lasers and also lock our free space filter cavities in front of the SNSPDs). The beating signal between these optical sidebands and the main carrier frequency now results in a coherent driving of the mechanical modes in a broad frequency range, which excites a mechanical packet inside the cavity. Following this excitation, we use a continuous red-detuned laser tone to continuously read the mechanical population. The resulting count rate from this read tone is shown in Fig.~\ref{Fig:coh_drive}a. We can clearly see the propagating behavior of the packet, first leaving the cavity resulting in a reduction of the population, followed by the reflection back to the cavity.

In order to verify that the reflection is indeed a result of the phonon leaving the cavity, we perform the same measurement on an identical device, with a 2.5 times longer waveguide attached to it. As expected, the longer waveguide should result in a 2.5 times longer traveling time of the packet, which is fully consistent with the time difference between the peaks in Fig.~\ref{Fig:coh_drive}b. Note that the two experiments are performed with different power settings of the coherent drive and the readout compared to the experiments in the main text, which is the reason of not seeing a full decay of the click rates in Fig.~\ref{Fig:coh_drive}a. This results in an increased thermal background introduced by the high power readout laser, as well as the pulse length of \SI{40}{ns} that is comparable with the round trip time of that device. Additionally, the peak widths are broader, since in this experiment only the modes within the bandwidth of the laser pulse will be excited. Therefore, the time domain behavior is limited by the bandwidth of the laser pulse as it is much narrower than the cavity-waveguide coupling and the bandwidth of the filter cavities.

\subsubsection{Scattering probabilities}

The Stokes and anti-Stokes scattering probabilities are determined from the click rates ($\Gamma\und{r}$, $\Gamma\und{b}$) on the SNSPDs and by measuring the detection efficiency of the experimental setup ($\eta\und{det}$)

\begin{eqnarray}
    \Gamma\und{r} = p\und{s,read}\cdot n\und{th}\cdot\eta\und{det}
    \label{eq:gamma_r}
    \\
    \Gamma\und{b} = p\und{s,write}\cdot(1+n\und{th})\cdot\eta\und{det}
    \label{eq:gamma_b}
\end{eqnarray}
    
where the $p\und{s,read}$ and $p\und{s,write}$ are the scattering probabilities from the red detuned (anti-Stokes) and blue detuned (Stokes) laser pulses, respectively, and $n\und{th}$ is the mechanical mode thermal phonon occupancy. The measurement of the detection efficiency gives $\eta\und{det} \approx 3.8 \%$. This includes the coupling between the optical cavity and the optical coupler waveguide $\eta\und{dev}$, the coupling from fiber to the optical coupler waveguide $\eta\und{c}$, the filter setup efficiency $\eta\und{F}$, the SNSPDs detection efficiency $\eta\und{SNSPD}$, as well as all other losses in the optical path $\eta\und{loss}$. Hence, the overall detection efficiency can be written as $\eta\und{det} = \eta\und{c}\cdot\eta\und{dev}\cdot\eta\und{F}\cdot\eta\und{SNSPD}\cdot\eta\und{loss}$. By shining a continuous off-resonance laser to the device and measuring the power at each point of the setup, we obtain $\eta\und{c} = 37 \%$, $\eta\und{F} = 38 \%$, $\eta\und{SNSPD} \approx 90 \%$ and $\eta\und{loss} \approx 80 \%$. Furthermore, the device efficiency $\eta\und{dev} = \frac{\kappa\und{e}}{\kappa\und{t}}$ and is measured using a similar method to~\cite{Groeblacher2013a}.

In the weak coupling regime ($g\ll\kappa$),  the scattering probabilities are given by the efficiency ($\eta\und{dev} = \frac{\kappa\und{e}}{\kappa\und{t}}$) and the incident laser pulse energy $E\und{p}$~\cite{Hong2017}:

\begin{eqnarray}
    p\und{s,read} = 1 - \exp\left(\frac{-4\eta\und{dev}g_{0}^{2}E\und{p}}{\hbar \omega\und{c}(\omega_{m}^{2} + (\kappa\und{t}/2)^{2})}\right)
    \\
    p\und{s,write} = \exp\left(\frac{4\eta\und{dev}g_{0}^{2}E\und{p}}{\hbar \omega\und{c}(\omega_{m}^{2} + (\kappa\und{t}/2)^{2})}\right) - 1
\end{eqnarray}

where $\omega\und{c}$ is the optical resonance frequency and $\omega_{m}$ the mechanical frequency.

\subsubsection{Band structure}

\begin{figure}[h!]
	\includegraphics[width = 1\linewidth]{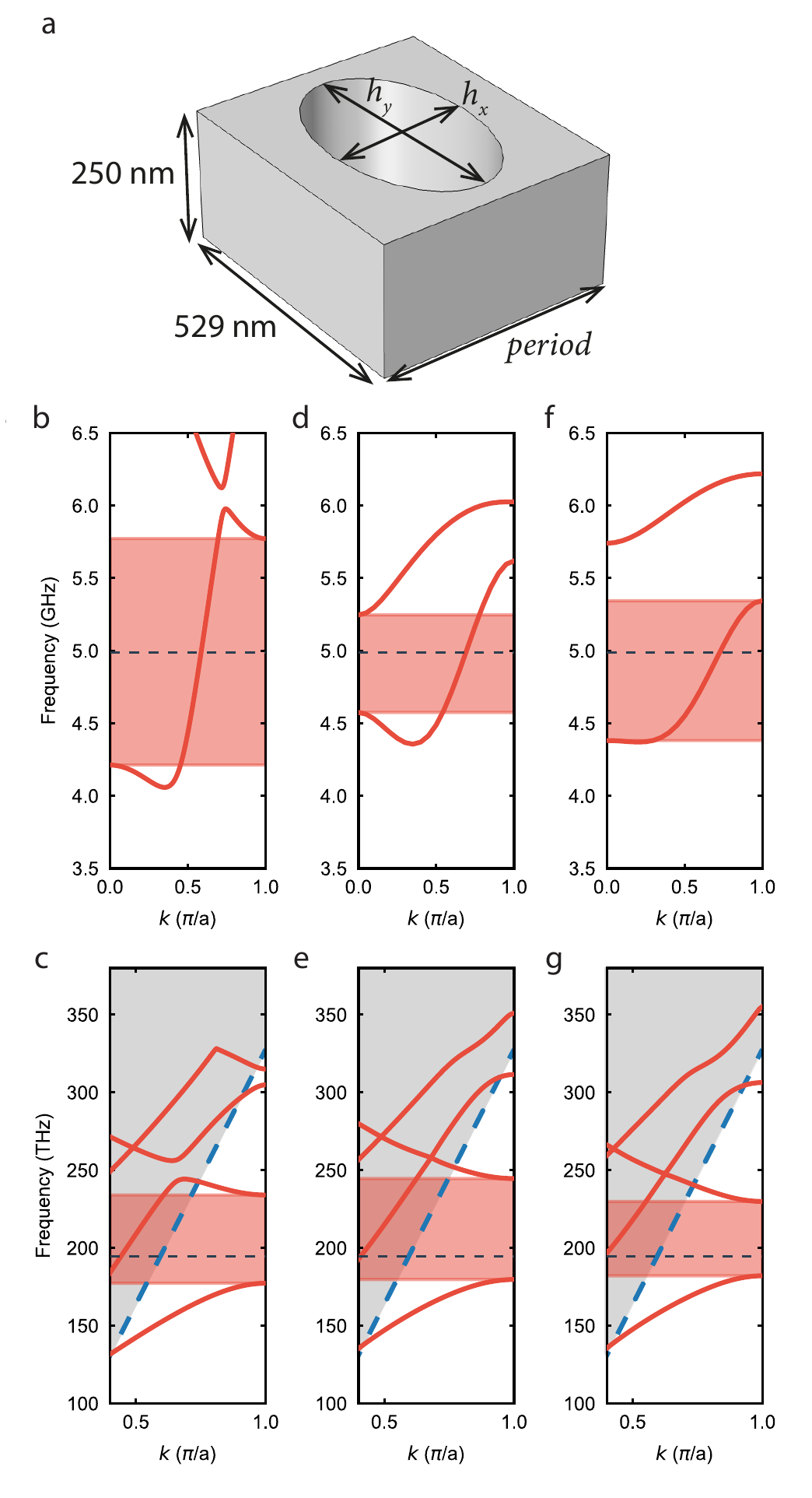}
	\caption{Engineering the band structure and phonon group velocity by adjusting the hole dimensions: $h\und{x}$ diameter along the waveguide orientation, $h\und{y}$, diameter perpendicular to the waveguide orientation, and the size of the unit cell (period). A sketch of the unit cell is shown in a). b,c) Mechanical and optical band diagram for a waveguide with parameters of $h\und{y} = \SI{187}{nm}$, $h\und{x} = \SI{320}{nm}$, $period = \SI{436}{nm}$ resulting in a mechanical group velocity of $v\und{g} = 1946~m/s$. d,e) Parameters of $h\und{y} = \SI{430}{nm}$, $h\und{x} = \SI{272}{nm}$, $period = \SI{458}{nm}$, with $v\und{g} = 2766~m/s$. f,g) Parameters of $h\und{y} = \SI{468}{nm}$, $hw = \SI{240}{nm}$, $period = \SI{458}{nm}$, with $v\und{g} = 6298~m/s$. In our experiment, we use the second set of parameters (shown in d) and e)) in order to have a small group velocity, while simultaneously maintaining a linear region of the mechanical band to minimize dispersion.}
	\label{Fig:S2}
\end{figure}

The band structure of the waveguide can be engineered by changing the hole dimensions and the periodicity of unit cells, while keeping the width and thickness of the beam fixed. In Fig.~\ref{Fig:S2} three designs of a waveguide with different mechanical band structure and group velocity are shown. All these designs act as optical mirrors for the cavity, as can be seen from the optical band diagrams.

\subsubsection{Lifetime}

\begin{figure}[h]
	\includegraphics[width = 1\linewidth]{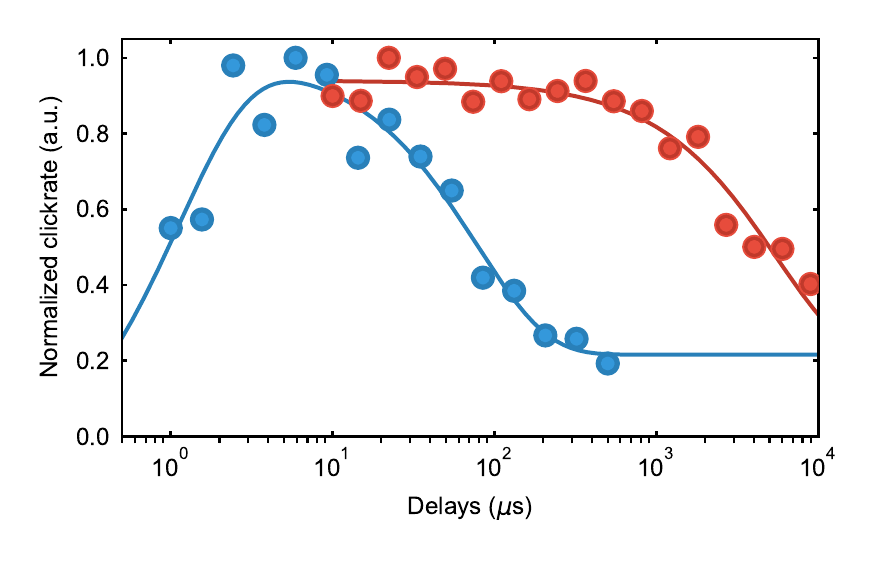}
	\caption{Normalized clickrates (directly proportional to the thermal population) from a thermally excited mechanical mode as a function of time. The population exhibits a double exponential behavior, which corresponds to the initial heating and subsequent decay. For the device used in the measurements the rise and decay times are $T\und{rise}\approx\SI{1}{\micro s}$ and $T_1\approx\SI{78}{\micro s}$ (data in blue), respectively. An exemplary second device, with increased number of phononic shield periods connected to the tethers, resulting in an much longer decay time $T_1\approx\SI{5.5}{ms}$ is also shown (data in red). The solid lines are exponential fits to the data.}
	\label{Fig:S3}
\end{figure}

One of the fundamental properties of the device is the extremely long lifetime $T_1$ of the mechanical excitation. In order to measure $T_1$ we use a red detuned strong optical pulse to heat the device, creating a thermal population in the mode. We then send another, much weaker, red detuned pulse, to probe the thermal population as a function of time, which is directly proportional to the clickrates. The result of the measurement is shown in Fig.~\ref{Fig:S3}. Note that, as previously seen~\cite{Wallucks2020}, the clickrates have a double exponential trend, with a rise time $T\und{rise}$ and a decay time $T_1$. For our measurement we choose one of the shorter lifetime devices (to allow for higher repetition rates), while several structures with up to \SI{5.5}{ms} are also fabricated on the same chip, as shown in the plot.

\subsubsection{Thermal occupancy of the mechanical mode}

\begin{figure}[ht!]
    \includegraphics[width = 1\linewidth]{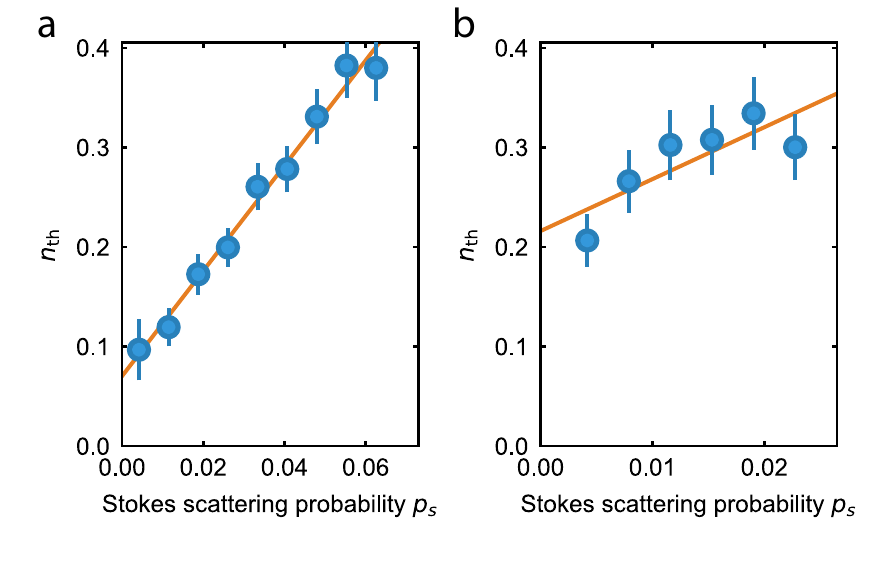}
    \caption{a) Thermal occupation of the mode of interest as a function of the scattering probability of the Stokes and anti-Stokes process $p\und{s,write} = p\und{s,read} = p\und{s}$. $n\und{th}$ is measured from the asymmetry of Stokes/anti-Stokes scattering rates. The fit follows $n\und{th} = (0.070 \pm 0.0095) + (5.29 \pm 0.25) \times p\und{s}$. We choose to work at $p\und{s,read} = 1.4\%$. b) Thermal occupation with the pulse scheme of the experiment as a function of the inferred scattering probability of the anti-Stokes process for the heating pulse. Here a heating pulse is sent to the device $\SI{170}{ns}$ before the pulses used to measure the thermal occupation. Since the heating dynamics is similar for both Stokes and anti-Stokes processes, we can send either a red-detuned or a blue-detuned pulse as heating pulse. The fitting of the resulting data follows $n\und{th} = (0.216 \pm 0.028) + (5.21 \pm 1.88) \times p\und{s}$.
    Choosing a scattering probability of $p\und{s,write} = 1.4\%$ results in a total thermal occupation of $n\und{th}\approx0.27$. In both figures the fits are linear and the errors are one standard deviation.}
    \label{Fig:S4}
\end{figure}

The mechanical modes, having a frequency of around $\SI{5}{GHz}$, have a thermal occupation of $<10^{-5}$ at $\SI{20}{mK}$. Due to absorption of the optical pulses used to create and read the state, the thermal occupation during the experiment is however significantly higher. We measure the mode temperature via the sideband asymmetry as shown in figure Fig.~\ref{Fig:S4}a.
In more detail, we send a blue-detuned and red-detuned optical pulse with exactly equal energies (and thus equal scattering probabilities) and a long delay between (few times of the mechanical lifetime in order to reinitialize to the ground state at the starting of each pulse) and measure the click rates from each pulse ($\Gamma\und{r}, \Gamma\und{b}$).
Using Eq.~\ref{eq:gamma_r} and \ref{eq:gamma_b}, we extract the thermal occupancy for different scattering probabilities.
This measurement allows us calibrate the mode heating that occurs during the optical red-detuned (anti-Stokes) pulse, the instantaneous heating.
To mimic the real experimental conditions with two pulses (as used for Fig.~\ref{Fig:3}a), we send an additional red detuned pulse to the device $\SI{170}{ns}$ before the pulses used to measure the thermal occupation, which effectively heats the mechanical mode. The results are shown in Fig.~\ref{Fig:S4}b, where the x-axis is the inferred scattering probability of the heating pulse.

\subsubsection{Error on the $g^{(2)}\und{om}$}

As discussed in the main text, we record the coincidences from the same trial ($N_{\Delta n = 0}$, where $n$ is an indicator for each experimental round), and the average number of coincidences from different trials ($\overline{N}_{\Delta n \neq 0}$), which is expected to be uncorrelated. In order to calculate the $g^{(2)}\und{om}$ we divide these two quantities, such that $g^{(2)}\und{om}(t) = N_{\Delta n = 0}(t) / \overline{N}_{\Delta n \neq 0}(t)$. Using $\Delta n\gg1$, one can gather enough statistics to estimate $\overline{N}_{\Delta n \neq 0}$ with a negligible error. For the estimation of the error on $N_{\Delta n = 0}$, which dominates the statistical uncertainty, a binomial distribution function is considered for the number of coincidences on each trial. Therefore, from this distribution, the probability of getting different numbers of coincidences can be calculated and thus, the error of the number of coincidences will be estimated. This is similar to the method used in~\cite{Riedinger2016}.

\subsubsection{Effect of time filtering window on $g^{(2)}\und{om}$}

\begin{figure}[h]
	\includegraphics[width = 1\linewidth]{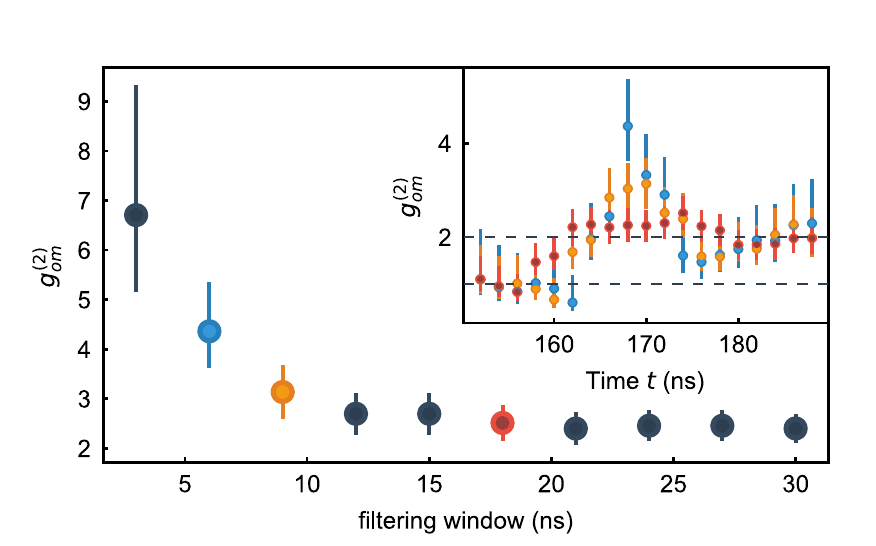}
	\caption{Maximum cross correlations obtained with varying filtering window width, using similar time delays as in Fig.~\ref{Fig:3}. Inset:\ full cross correlations as a function of time for 3 selected data points.}
	\label{Fig:gcc tw sweep}
\end{figure}

Due the strong coupling between the optomechanical defect and the mechanical waveguide, the phononic packet has a narrow length of around $T_{c} \approx  \SI{10}{ns}$. Therefore, in order to study the non-classical properties of the phononic packet, we have to filter the coincidences happening with a certain delay in time, with a narrow time window around that delay. In Fig.~\ref{Fig:3} the width of the time filtering window is \SI{6}{ns} and the delay of the window is scanned through the write and read pulses for each delay. The cross-correlation is calculated by dividing the number of coincidences happening in the same trial ($\Delta n = 0$), by the number of coincidences happening in different trials ($\Delta n \neq 0$), which are uncorrelated. By changing the filtering window, the number of uncorrelated coincidences resulting from the thermal background noise of mechanical motion is changed, reducing the cross correlations resulting from the heralded mechanical state. To further study this effect, we plot the cross correlations in Fig.~\ref{Fig:gcc tw sweep}, where we increase the filtering window from \SI{3}{ns} to \SI{30}{ns}. As can be seen, with a shorter filtering window the number of coincidences gathered for each delay is a reduced and thus, despite having a higher cross correlation, the error bars of the measurement is also larger. For window widths smaller than the packet lifetime we recover strong correlations, however for widths shorter than \SI{5}{ns} we only obtain a small number of coincidences, resulting in an increased uncertainty in the correlation parameter. For windows longer than the packet length, the increased contribution of uncorrelated coincidences reduces the measured correlations. Please note that the cross-correlations of classical state can never exceed 2.

\subsubsection{Double Pulse $g\und{om}^{(2)}(t)$}

\begin{figure*}[t]
	\includegraphics[width = 1\linewidth]{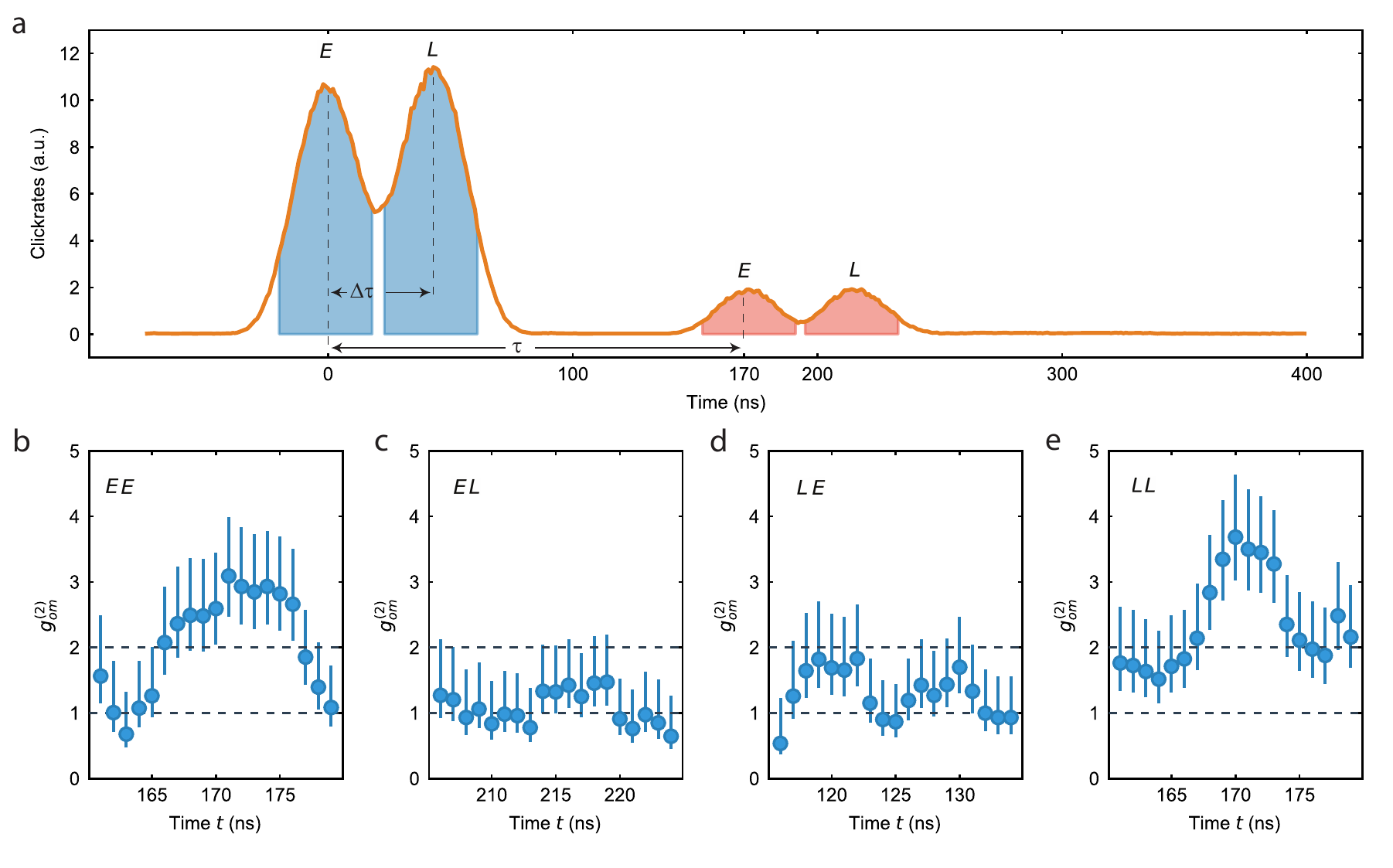}
	\caption{ a) Double-pulse click rate versus time. The two slightly overlapping write pulses preceding the read pulses are clearly distinguishable into earlier and later time. The delay between pulses is $\tau = \SI{170}{ns}$ for both "early-early" and "late-late" combinations, $\tau + \Delta \tau = \SI{215}{ns}$ for "early-late", and $\tau - \Delta \tau = \SI{125}{ns}$ for "late-early". b) Cross correlation between coincidences having a delay not more than $\SI{10}{ns}$ deviation from the delay between pulses, for different combinations. In all figures here "E"("L") stands for "early"("late").}
	\label{Fig:S5}
\end{figure*}

In this experiment we use two write and two read pulses. In Fig.~\ref{Fig:S5}a, the click rates are shown in time. The two pairs of pulses are clearly overlapping, since the delay between them is close to their FWHM. We gather coincidences from the shaded area of the pulses, choosing them to not overlap to not double count coincidences. We use a similar technique to the one use for Fig.~\ref{Fig:3}, with filtering the coincidences in a $\SI{6}{ns}$ time window with varying delays, in order to obtain finer time resolution. In Fig.~\ref{Fig:S5}b, the cross correlation between these fine-filtered coincidences is shown as a function of delay between them for different combinations of write and read pulse. The maximum cross correlation of these plots is depicted in Fig.~\ref{Fig:4}b. Note that the maximum cross correlation for "early-early" and "late-late" combinations happen at $t \approx \SI{170}{ns}$ which matches the second round trip of phonons, as expected.

\subsubsection{Theory of optomechanical induced transparency}

We assume to have an optomechanical cavity which is coupled to a truncated waveguide, acting as a Fabry-P\'{e}rot interferometer, and can write the total Hamiltonian ($\hbar = 1$) as

\begin{align*}
&H = \omega_{c}a^{\dagger}a + \omega_{m}b^{\dagger}b + \sum_{l}\omega_{l}c_{l}^{\dagger}c_{l} + \sum_{l}\gamma_{e,l}c_{l}^{\dagger}b + h.c.\\
&+ g_{0}a^{\dagger}a(b^{\dagger} + b)
\numberthis
\end{align*}

where $a$ and $b$ are the annihilation operators for the optical and mechanical modes of the optomechanical cavity and $g_{0}$ represents the single photon optomechanical coupling rate. Additionally, $\gamma_{e,l}$ represents the coupling between the mechanical mode of the cavity and the $l$-th mode of the Fabry-P\'{e}rot interferometer with annihilation operator $c_{l}$ and frequency $\omega_{l}$.

The linearized Hamiltonian in the rotating frame of the laser field after applying the perturbation over the optical field is then given by~\cite{Aspelmeyer2014}

\begin{align*}
&H = \Delta a^{\dagger}a + \omega_{m}b^{\dagger}b + g_{0}\alpha(a^{\dagger} + a)(b^{\dagger} + b)\\
& + \sum_{l} \Omega_{l}c_{l}^{\dagger}c_{l} + \sum_{l}\gamma_{e,l}c_{l}^{\dagger}b + h.c. \numberthis
\end{align*}

\begin{figure}[t]
	\includegraphics[width = 1\linewidth]{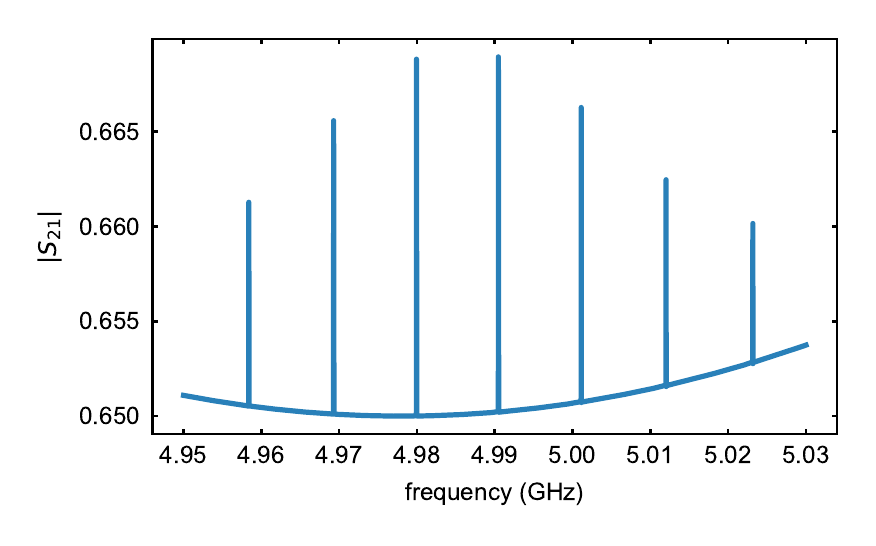}
	\caption{Numerically calculated $S_{21}$ signal, using the parameters given in the main text.}
	\label{Fig:S21_meas}
\end{figure}

where $\Delta = \omega_{c} - \omega_{l}$ is the detuning of the cavity with respect to the laser frequency. Here, $a$ is now the annihilation operator for the optical field fluctuation inside the cavity with the steady state field of $\alpha$. The equivalent optomechanical coupling to the field fluctuation is defined by $g = g_{0}.\alpha$.

The Langevin equations after applying the rotating wave approximation, by assuming having the pump field at the red sideband and neglecting the counter rotating terms, follow as

\begin{eqnarray}
	&\frac{da}{dt}& = -i\Delta a - igb - \frac{\kappa}{2} a +\sqrt{\kappa_{e}} a_{in} \label{eq:dadt}
	\\
	&\frac{db}{dt}& = -i\omega_{m}b - iga - \frac{\Gamma_{m}}{2}b -i\sum_{l}\gamma_{e,l}c_{l} \label{eq:dbdt}
	\\
	& \frac{dc_{l}}{dt}& = -i\omega_{l}c_{l} - i\gamma_{e,l}b - \frac{\Gamma_{l}}{2}b \label{eq:dcldt}
\end{eqnarray}

Here $\kappa$ is the optical decay which includes both external and internal cavity loss $\kappa = \kappa_{e} + \kappa_{i}$. Similarly, $\Gamma_{m}$ and $\Gamma_{l}$ are the mechanical decay rate for the cavity mechanical mode and $l$-th mechanical mode of the Fabry-P\'{e}rot interferometer, respectively. Also, $a_{in}$ are the cavity input field fluctuations.

For an OMIT measurement we pass the pump laser field through an amplitude EOM which produces sidebands at frequency $\omega$, thus the input field to the system can be written as $E_{0} + E_{+}e^{-i\omega t} + E_{-}e^{i\omega t}$. We are now interested in calculating the reflection coefficients of each component and eventually calculate the output field as $a_{out} = r_{0}E_{0} + r_{+}E_{+}e^{-i\omega t} + r_{-}E_{-}e^{i\omega t}$. Without loss of generality, we can assume that the two sidebands are equal and produced in-phase ($E_{+} = E_{-} = E_{1}$) by the amplitude EOM. Moreover, by operating in the resolved sideband regime ($\kappa < \omega_{m},\Delta$), the fields at frequencies far detuned from the cavity frequency reflect identically and we can assume $r_{0} \approx r_{-} \approx 1$. We therefore have to only calculate the reflection coefficient of the component close to the cavity resonance $r_{+}$. By considering only this input field in Eq.~\ref{eq:dadt} and assuming $a = a_{+}e^{-i\omega t}$, $b = b_{+}e^{-i\omega t}$ and $c_{l} = c_{+,l}e^{-i\omega t}$, we have

\begin{figure}[b]
	\includegraphics[width = 1\linewidth]{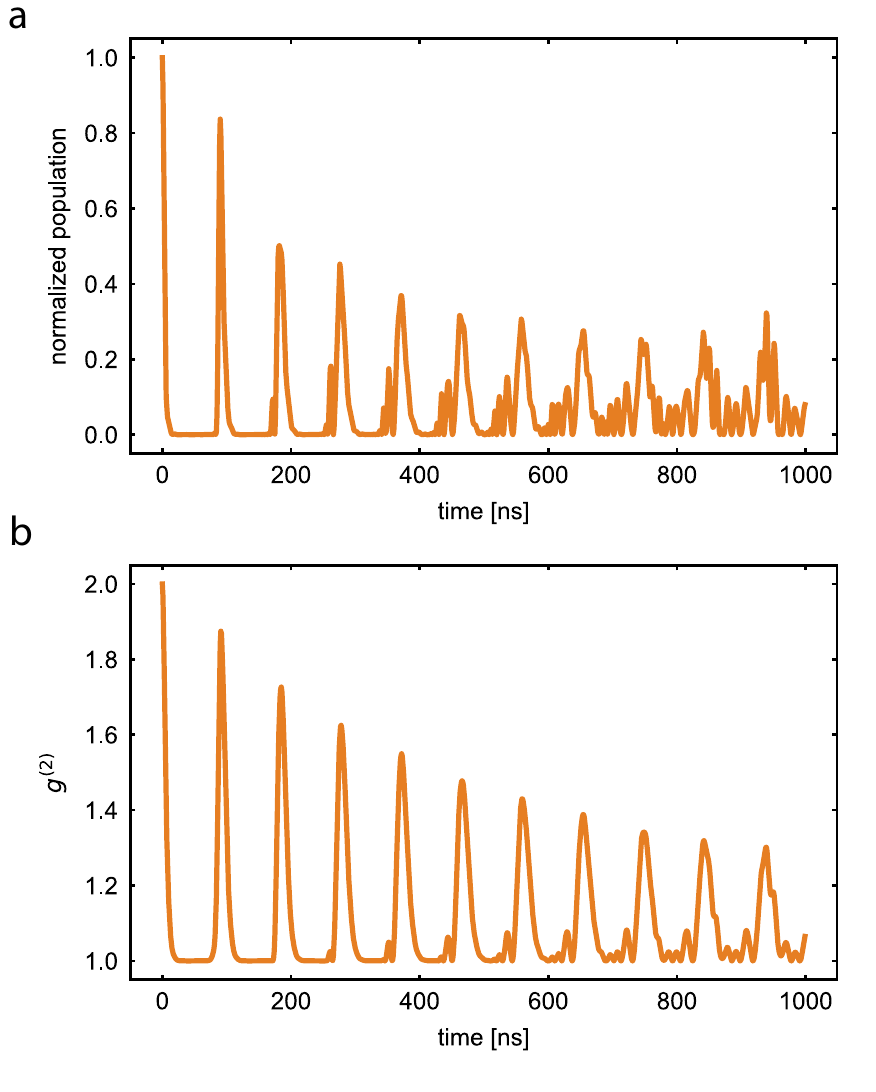}
	\caption{a) Simulated normalized cavity population as a function of time, detected after passing through the filter cavities. b) Second order correlation function of the cavity mechanical field, calculated using the Siegert relation.}
	\label{Fig:ng2_v3}
\end{figure}

\begin{eqnarray}
	&&b_{+} = \frac{g}{\omega - \omega_{m} + i\frac{\Gamma_{m}}{2} - \sum_{l}\frac{\gamma_{e,l}^{2}}{\omega - \omega_{l} + i\frac{\Gamma_{l}}{2}}} a_{+}
	\\
	&&a_{+} = \frac{i\kappa_{e}}{\omega - \Delta + i\frac{\kappa}{2} - \chi(\omega)} E_{1}
	\\
	&& \chi(\omega) = \frac{g^{2}}{\omega - \omega_{m} + i\frac{\Gamma_{m}}{2} - \sum_{l}\frac{\gamma_{e,l}^{2}}{\omega - \omega_{l} + i\frac{\Gamma_{l}}{2}}}
\end{eqnarray}

with $\chi(\omega)$ being the mechanical susceptibility to the optical field. After that, by using the input-output relation of $a_{out} = a_{in} - \sqrt{\kappa_{e}} a$, we can extract the reflection coefficient as

\begin{equation}
r_{+} = 1 - \frac{i\kappa_{e}}{\omega - \Delta + i\frac{\kappa}{2} - \chi(\omega)} \label{eq:r}
\end{equation}

A fast photodiode detects the power of the output field as

\begin{align*}
P = |E_{0}|^{2} + |E_{1}|^{2} + |r_{+}E_{1}|^2 + E_{0}E_{1}e^{-i\omega t}+
\\
r_{+}E_{0}E_{1}e^{-i\omega t} + r_{+}|E_{1}|^2e^{-i2\omega t} + c.c. \label{eq:Pout} \numberthis
\end{align*}

The voltage produced by the fast photodiode is connected to channel 2 of a vector network analyzer (VNA) which measures the quadratures of the signal, and thus measures $S_{21}$ with magnitude of $|S_{21}| = \frac{1 + Re\{r_{+}\}}{2}$. For simplicity, we assume all the Fabry-P\'{e}rot interferometer modes to have a similar coupling rate to the cavity. We now numerically calculate and plot this signal in Fig.~\ref{Fig:S21_meas} given the parameters of the system, which are reported in the main text. The resulting graph is very similar to the measurement shown in Fig.~\ref{Fig:2}b. Note that the peaks in the transparency window at each mechanical mode are a direct result of the optomechanical coupling in the denominator of Eq.~\ref{eq:r}.

\begin{figure}[ht]
	\includegraphics[width = 1\linewidth]{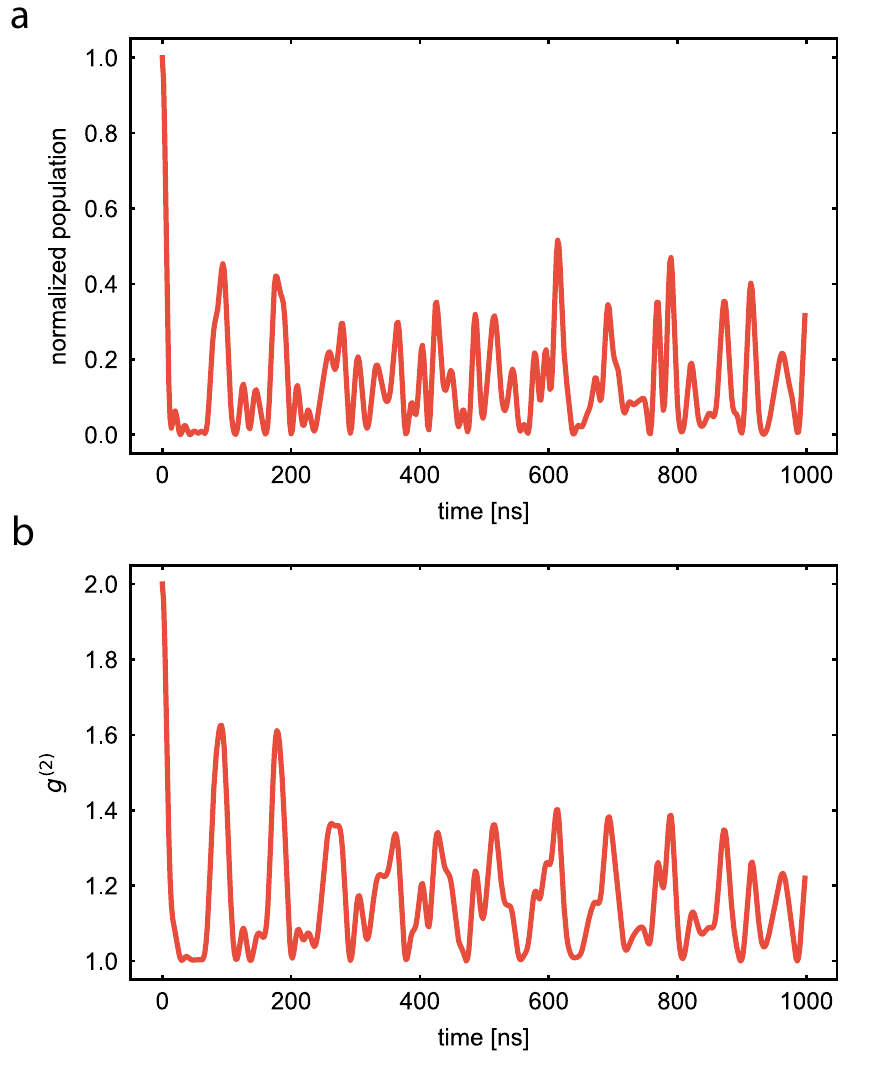}
	\caption{We apply the same model to the OMIT spectrum measured for the real device in the main text, obtaining the cavity population and $g^{(2)}$ of a thermal state as a function of time.}
	\label{Fig:ng2_omit_v3}
\end{figure}

\subsubsection{Mechanical response of the cavity-waveguide structure}

In order to develop a simple theoretical model describing the measurements in Fig.~\ref{Fig:2}c in the main text we first look at the time domain behavior of the mechanical system and then calculate its $g^{(2)}$ function. The state-swap interaction performed with the red-detuned optical drive is such that the optical $g^{(2)}$ we recover reproduces the mechanical state~\cite{Hong2017}. For the time domain response we use Eq.~\ref{eq:dbdt} and Eq.~\ref{eq:dcldt}, where we set the optomechanical interaction term to zero ($g\approx 0$). We solve them using a Python module that solves a system of linearly coupled differential equations (Scipy.integrate.odeint module), with the initial conditions of having a unity population in the cavity ($b(0) = 1$) and all the Fabry-P\'{e}rot interferometer modes in the vacuum state ($c_{l}(0) = 0$). For simplicity, we solve the time dynamics for scalar classical fields. We then use the solution of $b(t)$ to further obtain the cavity population and the first order correlation function $g^{(1)}$

\begin{equation}
g^{(1)}_\tau = \frac{\int\left\langle b^{\dagger}(t)b(t + \tau)\right\rangle dt}{\int \left\langle|b(t)|^{2}\right\rangle dt}
\end{equation}

We then proceed by using the Siegert relation~\cite{Lebreton2013} to calculate the second order correlation $g^{(2)}_\tau$ of a thermal chaotic field

\begin{equation}
g^{(2)}_\tau = 1 + |g^{(1)}_\tau|^{2}
\end{equation}

The general results for our simple model including the time domain response of the cavity population, as well as the intensity correlation $g^{(2)}_\tau$ are shown in Fig.~\ref{Fig:ng2_v3}. The model clearly shows that having the adjacent waveguide results in a leakage of the phonon population, followed by a subsequent revival after each round-trip with the reflection from the free-standing end. A similar behavior is observed for the intensity correlation, where $g^{(2)}_\tau$ decays as the thermal field leaves the cavity and then rises again as the mechanical field reflects back into the cavity.

We can now use the same method in combination with the measured OMIT data as the input and simulate the expected time domain response, as well as the mechanical $g^{(2)}_\tau$ as a function of time, using the same assumptions as above (see Fig.~\ref{Fig:ng2_omit_v3}). This simple model allows us to qualitatively model our measured data from Fig.~\ref{Fig:2}c very well.


\begin{thebibliography}{35}%
\makeatletter
\providecommand \@ifxundefined [1]{%
 \@ifx{#1\undefined}
}%
\providecommand \@ifnum [1]{%
 \ifnum #1\expandafter \@firstoftwo
 \else \expandafter \@secondoftwo
 \fi
}%
\providecommand \@ifx [1]{%
 \ifx #1\expandafter \@firstoftwo
 \else \expandafter \@secondoftwo
 \fi
}%
\providecommand \natexlab [1]{#1}%
\providecommand \enquote  [1]{``#1''}%
\providecommand \bibnamefont  [1]{#1}%
\providecommand \bibfnamefont [1]{#1}%
\providecommand \citenamefont [1]{#1}%
\providecommand \href@noop [0]{\@secondoftwo}%
\providecommand \href [0]{\begingroup \@sanitize@url \@href}%
\providecommand \@href[1]{\@@startlink{#1}\@@href}%
\providecommand \@@href[1]{\endgroup#1\@@endlink}%
\providecommand \@sanitize@url [0]{\catcode `\\12\catcode `\$12\catcode
  `\&12\catcode `\#12\catcode `\^12\catcode `\_12\catcode `\%12\relax}%
\providecommand \@@startlink[1]{}%
\providecommand \@@endlink[0]{}%
\providecommand \url  [0]{\begingroup\@sanitize@url \@url }%
\providecommand \@url [1]{\endgroup\@href {#1}{\urlprefix }}%
\providecommand \urlprefix  [0]{URL }%
\providecommand \Eprint [0]{\href }%
\providecommand \doibase [0]{https://doi.org/}%
\providecommand \selectlanguage [0]{\@gobble}%
\providecommand \bibinfo  [0]{\@secondoftwo}%
\providecommand \bibfield  [0]{\@secondoftwo}%
\providecommand \translation [1]{[#1]}%
\providecommand \BibitemOpen [0]{}%
\providecommand \bibitemStop [0]{}%
\providecommand \bibitemNoStop [0]{.\EOS\space}%
\providecommand \EOS [0]{\spacefactor3000\relax}%
\providecommand \BibitemShut  [1]{\csname bibitem#1\endcsname}%
\let\auto@bib@innerbib\@empty
\bibitem [{\citenamefont {Aspelmeyer}\ \emph {et~al.}(2014)\citenamefont
  {Aspelmeyer}, \citenamefont {Kippenberg},\ and\ \citenamefont
  {Marquardt}}]{Aspelmeyer2014}%
  \BibitemOpen
  \bibfield  {author} {\bibinfo {author} {\bibfnamefont {M.}~\bibnamefont
  {Aspelmeyer}}, \bibinfo {author} {\bibfnamefont {T.~J.}\ \bibnamefont
  {Kippenberg}},\ and\ \bibinfo {author} {\bibfnamefont {F.}~\bibnamefont
  {Marquardt}},\ }\bibfield  {title} {\bibinfo {title} {Cavity optomechanics},\
  }\href {https://doi.org/10.1103/RevModPhys.86.1391} {\bibfield  {journal}
  {\bibinfo  {journal} {Rev.\ Mod.\ Phys.}\ }\textbf {\bibinfo {volume} {86}},\
  \bibinfo {pages} {1391} (\bibinfo {year} {2014})}\BibitemShut {NoStop}%
\bibitem [{\citenamefont {O'Connell}\ \emph {et~al.}(2010)\citenamefont
  {O'Connell}, \citenamefont {Hofheinz}, \citenamefont {Ansmann}, \citenamefont
  {Bialczak}, \citenamefont {Lenander}, \citenamefont {Lucero}, \citenamefont
  {Neeley}, \citenamefont {Sank}, \citenamefont {Wang}, \citenamefont {Weides},
  \citenamefont {Wenner}, \citenamefont {Martinis},\ and\ \citenamefont
  {Cleland}}]{OConnell2010}%
  \BibitemOpen
  \bibfield  {author} {\bibinfo {author} {\bibfnamefont {A.~D.}\ \bibnamefont
  {O'Connell}}, \bibinfo {author} {\bibfnamefont {M.}~\bibnamefont {Hofheinz}},
  \bibinfo {author} {\bibfnamefont {M.}~\bibnamefont {Ansmann}}, \bibinfo
  {author} {\bibfnamefont {R.~C.}\ \bibnamefont {Bialczak}}, \bibinfo {author}
  {\bibfnamefont {M.}~\bibnamefont {Lenander}}, \bibinfo {author}
  {\bibfnamefont {E.}~\bibnamefont {Lucero}}, \bibinfo {author} {\bibfnamefont
  {M.}~\bibnamefont {Neeley}}, \bibinfo {author} {\bibfnamefont
  {D.}~\bibnamefont {Sank}}, \bibinfo {author} {\bibfnamefont {H.}~\bibnamefont
  {Wang}}, \bibinfo {author} {\bibfnamefont {M.}~\bibnamefont {Weides}},
  \bibinfo {author} {\bibfnamefont {J.}~\bibnamefont {Wenner}}, \bibinfo
  {author} {\bibfnamefont {J.~M.}\ \bibnamefont {Martinis}},\ and\ \bibinfo
  {author} {\bibfnamefont {A.~N.}\ \bibnamefont {Cleland}},\ }\bibfield
  {title} {\bibinfo {title} {Quantum ground state and single-phonon control of
  a mechanical resonator},\ }\href {https://doi.org/10.1038/nature08967}
  {\bibfield  {journal} {\bibinfo  {journal} {Nature}\ }\textbf {\bibinfo
  {volume} {464}},\ \bibinfo {pages} {697} (\bibinfo {year}
  {2010})}\BibitemShut {NoStop}%
\bibitem [{\citenamefont {Palomaki}\ \emph {et~al.}(2013)\citenamefont
  {Palomaki}, \citenamefont {Teufel}, \citenamefont {Simmonds},\ and\
  \citenamefont {Lehnert}}]{Palomaki2013}%
  \BibitemOpen
  \bibfield  {author} {\bibinfo {author} {\bibfnamefont {T.}~\bibnamefont
  {Palomaki}}, \bibinfo {author} {\bibfnamefont {J.}~\bibnamefont {Teufel}},
  \bibinfo {author} {\bibfnamefont {R.}~\bibnamefont {Simmonds}},\ and\
  \bibinfo {author} {\bibfnamefont {K.}~\bibnamefont {Lehnert}},\ }\bibfield
  {title} {\bibinfo {title} {Entangling mechanical motion with microwave
  fields},\ }\href {https://doi.org/10.1126/science.1244563} {\bibfield
  {journal} {\bibinfo  {journal} {Science}\ }\textbf {\bibinfo {volume}
  {342}},\ \bibinfo {pages} {710} (\bibinfo {year} {2013})}\BibitemShut
  {NoStop}%
\bibitem [{\citenamefont {Riedinger}\ \emph {et~al.}(2018)\citenamefont
  {Riedinger}, \citenamefont {Wallucks}, \citenamefont {Marinkovi\'{c}},
  \citenamefont {L\"oschnauer}, \citenamefont {Aspelmeyer}, \citenamefont
  {Hong},\ and\ \citenamefont {Gr\"oblacher}}]{Riedinger2018}%
  \BibitemOpen
  \bibfield  {author} {\bibinfo {author} {\bibfnamefont {R.}~\bibnamefont
  {Riedinger}}, \bibinfo {author} {\bibfnamefont {A.}~\bibnamefont {Wallucks}},
  \bibinfo {author} {\bibfnamefont {I.}~\bibnamefont {Marinkovi\'{c}}},
  \bibinfo {author} {\bibfnamefont {C.}~\bibnamefont {L\"oschnauer}}, \bibinfo
  {author} {\bibfnamefont {M.}~\bibnamefont {Aspelmeyer}}, \bibinfo {author}
  {\bibfnamefont {S.}~\bibnamefont {Hong}},\ and\ \bibinfo {author}
  {\bibfnamefont {S.}~\bibnamefont {Gr\"oblacher}},\ }\bibfield  {title}
  {\bibinfo {title} {Remote quantum entanglement between two micromechanical
  oscillators},\ }\href {https://doi.org/10.1038/s41586-018-0036-z} {\bibfield
  {journal} {\bibinfo  {journal} {Nature}\ }\textbf {\bibinfo {volume} {556}},\
  \bibinfo {pages} {473} (\bibinfo {year} {2018})}\BibitemShut {NoStop}%
\bibitem [{\citenamefont {Ockeloen-Korppi}\ \emph {et~al.}(2018)\citenamefont
  {Ockeloen-Korppi}, \citenamefont {Damsk\"agg}, \citenamefont {Pirkkalainen},
  \citenamefont {Asjad}, \citenamefont {Clerk}, \citenamefont {Massel},
  \citenamefont {Woolley},\ and\ \citenamefont
  {Sillanp\"a\"a}}]{Ockeloen-Korppi2018}%
  \BibitemOpen
  \bibfield  {author} {\bibinfo {author} {\bibfnamefont {C.~F.}\ \bibnamefont
  {Ockeloen-Korppi}}, \bibinfo {author} {\bibfnamefont {E.}~\bibnamefont
  {Damsk\"agg}}, \bibinfo {author} {\bibfnamefont {J.-M.}\ \bibnamefont
  {Pirkkalainen}}, \bibinfo {author} {\bibfnamefont {M.}~\bibnamefont {Asjad}},
  \bibinfo {author} {\bibfnamefont {A.~A.}\ \bibnamefont {Clerk}}, \bibinfo
  {author} {\bibfnamefont {F.}~\bibnamefont {Massel}}, \bibinfo {author}
  {\bibfnamefont {M.~J.}\ \bibnamefont {Woolley}},\ and\ \bibinfo {author}
  {\bibfnamefont {M.~A.}\ \bibnamefont {Sillanp\"a\"a}},\ }\bibfield  {title}
  {\bibinfo {title} {Stabilized entanglement of massive mechanical
  oscillators},\ }\href {https://doi.org/10.1038/s41586-018-0038-x} {\bibfield
  {journal} {\bibinfo  {journal} {Nature}\ }\textbf {\bibinfo {volume} {556}},\
  \bibinfo {pages} {478} (\bibinfo {year} {2018})}\BibitemShut {NoStop}%
\bibitem [{\citenamefont {Marinkovi\'{c}}\ \emph {et~al.}(2018)\citenamefont
  {Marinkovi\'{c}}, \citenamefont {Wallucks}, \citenamefont {Riedinger},
  \citenamefont {Hong}, \citenamefont {Aspelmeyer},\ and\ \citenamefont
  {Gr\"oblacher}}]{Marinkovic2018}%
  \BibitemOpen
  \bibfield  {author} {\bibinfo {author} {\bibfnamefont {I.}~\bibnamefont
  {Marinkovi\'{c}}}, \bibinfo {author} {\bibfnamefont {A.}~\bibnamefont
  {Wallucks}}, \bibinfo {author} {\bibfnamefont {R.}~\bibnamefont {Riedinger}},
  \bibinfo {author} {\bibfnamefont {S.}~\bibnamefont {Hong}}, \bibinfo {author}
  {\bibfnamefont {M.}~\bibnamefont {Aspelmeyer}},\ and\ \bibinfo {author}
  {\bibfnamefont {S.}~\bibnamefont {Gr\"oblacher}},\ }\bibfield  {title}
  {\bibinfo {title} {{An optomechanical Bell test}},\ }\href
  {https://doi.org/10.1103/PhysRevLett.121.220404} {\bibfield  {journal}
  {\bibinfo  {journal} {Phys.\ Rev.\ Lett.}\ }\textbf {\bibinfo {volume}
  {121}},\ \bibinfo {pages} {220404} (\bibinfo {year} {2018})}\BibitemShut
  {NoStop}%
\bibitem [{\citenamefont {Wallucks}\ \emph {et~al.}(2020)\citenamefont
  {Wallucks}, \citenamefont {Marinkovi\'{c}}, \citenamefont {Hensen},
  \citenamefont {Stockill},\ and\ \citenamefont {Gr\"oblacher}}]{Wallucks2020}%
  \BibitemOpen
  \bibfield  {author} {\bibinfo {author} {\bibfnamefont {A.}~\bibnamefont
  {Wallucks}}, \bibinfo {author} {\bibfnamefont {I.}~\bibnamefont
  {Marinkovi\'{c}}}, \bibinfo {author} {\bibfnamefont {B.}~\bibnamefont
  {Hensen}}, \bibinfo {author} {\bibfnamefont {R.}~\bibnamefont {Stockill}},\
  and\ \bibinfo {author} {\bibfnamefont {S.}~\bibnamefont {Gr\"oblacher}},\
  }\bibfield  {title} {\bibinfo {title} {A quantum memory at telecom
  wavelengths},\ }\href {https://doi.org/10.1038/s41567-020-0891-z} {\bibfield
  {journal} {\bibinfo  {journal} {Nature Phys.}\ }\textbf {\bibinfo {volume}
  {16}},\ \bibinfo {pages} {772} (\bibinfo {year} {2020})}\BibitemShut
  {NoStop}%
\bibitem [{\citenamefont {Andrews}\ \emph {et~al.}(2014)\citenamefont
  {Andrews}, \citenamefont {Peterson}, \citenamefont {Purdy}, \citenamefont
  {Cicak}, \citenamefont {Simmonds}, \citenamefont {Regal},\ and\ \citenamefont
  {Lehnert}}]{Andrews2014}%
  \BibitemOpen
  \bibfield  {author} {\bibinfo {author} {\bibfnamefont {R.~W.}\ \bibnamefont
  {Andrews}}, \bibinfo {author} {\bibfnamefont {R.~W.}\ \bibnamefont
  {Peterson}}, \bibinfo {author} {\bibfnamefont {T.~P.}\ \bibnamefont {Purdy}},
  \bibinfo {author} {\bibfnamefont {K.}~\bibnamefont {Cicak}}, \bibinfo
  {author} {\bibfnamefont {R.~W.}\ \bibnamefont {Simmonds}}, \bibinfo {author}
  {\bibfnamefont {C.~A.}\ \bibnamefont {Regal}},\ and\ \bibinfo {author}
  {\bibfnamefont {K.~W.}\ \bibnamefont {Lehnert}},\ }\bibfield  {title}
  {\bibinfo {title} {{Bidirectional and efficient conversion between microwave
  and optical light}},\ }\href {https://doi.org/10.1038/nphys2911} {\bibfield
  {journal} {\bibinfo  {journal} {Nature Phys.}\ }\textbf {\bibinfo {volume}
  {10}},\ \bibinfo {pages} {321} (\bibinfo {year} {2014})}\BibitemShut
  {NoStop}%
\bibitem [{\citenamefont {Vainsencher}\ \emph {et~al.}(2016)\citenamefont
  {Vainsencher}, \citenamefont {Satzinger}, \citenamefont {Peairs},\ and\
  \citenamefont {Cleland}}]{Vainsencher2016}%
  \BibitemOpen
  \bibfield  {author} {\bibinfo {author} {\bibfnamefont {A.}~\bibnamefont
  {Vainsencher}}, \bibinfo {author} {\bibfnamefont {K.~J.}\ \bibnamefont
  {Satzinger}}, \bibinfo {author} {\bibfnamefont {G.~A.}\ \bibnamefont
  {Peairs}},\ and\ \bibinfo {author} {\bibfnamefont {A.~N.}\ \bibnamefont
  {Cleland}},\ }\bibfield  {title} {\bibinfo {title} {Bi-directional conversion
  between microwave and optical frequencies in a piezoelectric optomechanical
  device},\ }\href {https://doi.org/10.1063/1.4955408} {\bibfield  {journal}
  {\bibinfo  {journal} {Appl.\ Phys.\ Lett.}\ }\textbf {\bibinfo {volume}
  {109}},\ \bibinfo {pages} {033107} (\bibinfo {year} {2016})}\BibitemShut
  {NoStop}%
\bibitem [{\citenamefont {Forsch}\ \emph {et~al.}(2020)\citenamefont {Forsch},
  \citenamefont {Stockill}, \citenamefont {Wallucks}, \citenamefont
  {Marinkovi\'{c}}, \citenamefont {G\"{a}rtner}, \citenamefont {Norte},
  \citenamefont {van Otten}, \citenamefont {Fiore}, \citenamefont
  {Srinivasan},\ and\ \citenamefont {Gr\"{o}blacher}}]{Forsch2020}%
  \BibitemOpen
  \bibfield  {author} {\bibinfo {author} {\bibfnamefont {M.}~\bibnamefont
  {Forsch}}, \bibinfo {author} {\bibfnamefont {R.}~\bibnamefont {Stockill}},
  \bibinfo {author} {\bibfnamefont {A.}~\bibnamefont {Wallucks}}, \bibinfo
  {author} {\bibfnamefont {I.}~\bibnamefont {Marinkovi\'{c}}}, \bibinfo
  {author} {\bibfnamefont {C.}~\bibnamefont {G\"{a}rtner}}, \bibinfo {author}
  {\bibfnamefont {R.~A.}\ \bibnamefont {Norte}}, \bibinfo {author}
  {\bibfnamefont {F.}~\bibnamefont {van Otten}}, \bibinfo {author}
  {\bibfnamefont {A.}~\bibnamefont {Fiore}}, \bibinfo {author} {\bibfnamefont
  {K.}~\bibnamefont {Srinivasan}},\ and\ \bibinfo {author} {\bibfnamefont
  {S.}~\bibnamefont {Gr\"{o}blacher}},\ }\bibfield  {title} {\bibinfo {title}
  {Microwave-to-optics conversion using a mechanical oscillator in its quantum
  groundstate},\ }\href {https://doi.org/10.1038/s41567-019-0673-7} {\bibfield
  {journal} {\bibinfo  {journal} {Nature Phys.}\ }\textbf {\bibinfo {volume}
  {16}},\ \bibinfo {pages} {69} (\bibinfo {year} {2020})}\BibitemShut {NoStop}%
\bibitem [{\citenamefont {Mirhosseini}\ \emph {et~al.}(2020)\citenamefont
  {Mirhosseini}, \citenamefont {Sipahigil}, \citenamefont {Kalaee},\ and\
  \citenamefont {Painter}}]{Mirhosseini2020}%
  \BibitemOpen
  \bibfield  {author} {\bibinfo {author} {\bibfnamefont {M.}~\bibnamefont
  {Mirhosseini}}, \bibinfo {author} {\bibfnamefont {A.}~\bibnamefont
  {Sipahigil}}, \bibinfo {author} {\bibfnamefont {M.}~\bibnamefont {Kalaee}},\
  and\ \bibinfo {author} {\bibfnamefont {O.}~\bibnamefont {Painter}},\
  }\bibfield  {title} {\bibinfo {title} {Superconducting qubit to optical
  photon transduction},\ }\href {https://doi.org/10.1038/s41586-020-3038-6}
  {\bibfield  {journal} {\bibinfo  {journal} {Nature}\ }\textbf {\bibinfo
  {volume} {588}},\ \bibinfo {pages} {599} (\bibinfo {year}
  {2020})}\BibitemShut {NoStop}%
\bibitem [{\citenamefont {Habraken}\ \emph {et~al.}(2012)\citenamefont
  {Habraken}, \citenamefont {Stannigel}, \citenamefont {Lukin}, \citenamefont
  {Zoller},\ and\ \citenamefont {Rabl}}]{Habraken2012}%
  \BibitemOpen
  \bibfield  {author} {\bibinfo {author} {\bibfnamefont {S.~J.~M.}\
  \bibnamefont {Habraken}}, \bibinfo {author} {\bibfnamefont {K.}~\bibnamefont
  {Stannigel}}, \bibinfo {author} {\bibfnamefont {M.~D.}\ \bibnamefont
  {Lukin}}, \bibinfo {author} {\bibfnamefont {P.}~\bibnamefont {Zoller}},\ and\
  \bibinfo {author} {\bibfnamefont {P.}~\bibnamefont {Rabl}},\ }\bibfield
  {title} {\bibinfo {title} {{Continuous mode cooling and phonon routers for
  phononic quantum networks}},\ }\href
  {https://doi.org/10.1088/1367-2630/14/11/115004} {\bibfield  {journal}
  {\bibinfo  {journal} {New J.\ Phys.}\ }\textbf {\bibinfo {volume} {14}},\
  \bibinfo {pages} {115004} (\bibinfo {year} {2012})}\BibitemShut {NoStop}%
\bibitem [{\citenamefont {Delsing}\ \emph {et~al.}(2019)\citenamefont {Delsing}
  \emph {et~al.}}]{Delsing2019}%
  \BibitemOpen
  \bibfield  {author} {\bibinfo {author} {\bibfnamefont {P.}~\bibnamefont
  {Delsing}} \emph {et~al.},\ }\bibfield  {title} {\bibinfo {title} {The 2019
  surface acoustic waves roadmap},\ }\href
  {https://doi.org/10.1088/1361-6463/ab1b04} {\bibfield  {journal} {\bibinfo
  {journal} {J.\ Phys.\ D:\ Appl.\ Phys.}\ }\textbf {\bibinfo {volume} {52}},\
  \bibinfo {pages} {353001} (\bibinfo {year} {2019})}\BibitemShut {NoStop}%
\bibitem [{\citenamefont {Schuetz}\ \emph {et~al.}(2015)\citenamefont
  {Schuetz}, \citenamefont {Kessler}, \citenamefont {Giedke}, \citenamefont
  {Vandersypen}, \citenamefont {Lukin},\ and\ \citenamefont
  {Cirac}}]{Schuetz2015}%
  \BibitemOpen
  \bibfield  {author} {\bibinfo {author} {\bibfnamefont {M.~J.~A.}\
  \bibnamefont {Schuetz}}, \bibinfo {author} {\bibfnamefont {E.~M.}\
  \bibnamefont {Kessler}}, \bibinfo {author} {\bibfnamefont {G.}~\bibnamefont
  {Giedke}}, \bibinfo {author} {\bibfnamefont {L.~M.~K.}\ \bibnamefont
  {Vandersypen}}, \bibinfo {author} {\bibfnamefont {M.~D.}\ \bibnamefont
  {Lukin}},\ and\ \bibinfo {author} {\bibfnamefont {J.~I.}\ \bibnamefont
  {Cirac}},\ }\bibfield  {title} {\bibinfo {title} {{Universal Quantum
  Transducers Based on Surface Acoustic Waves}},\ }\href
  {https://doi.org/10.1103/PhysRevX.5.031031} {\bibfield  {journal} {\bibinfo
  {journal} {Phys.\ Rev.\ X}\ }\textbf {\bibinfo {volume} {5}},\ \bibinfo
  {pages} {031031} (\bibinfo {year} {2015})}\BibitemShut {NoStop}%
\bibitem [{\citenamefont {Bienfait}\ \emph {et~al.}(2019)\citenamefont
  {Bienfait}, \citenamefont {Satzinger}, \citenamefont {Zhong}, \citenamefont
  {Chang}, \citenamefont {Chou}, \citenamefont {Conner}, \citenamefont {Dumur},
  \citenamefont {Grebel}, \citenamefont {Peairs}, \citenamefont {Povey},\ and\
  \citenamefont {Cleland}}]{Bienfait2019}%
  \BibitemOpen
  \bibfield  {author} {\bibinfo {author} {\bibfnamefont {A.}~\bibnamefont
  {Bienfait}}, \bibinfo {author} {\bibfnamefont {K.~J.}\ \bibnamefont
  {Satzinger}}, \bibinfo {author} {\bibfnamefont {Y.~P.}\ \bibnamefont
  {Zhong}}, \bibinfo {author} {\bibfnamefont {H.-S.}\ \bibnamefont {Chang}},
  \bibinfo {author} {\bibfnamefont {M.-H.}\ \bibnamefont {Chou}}, \bibinfo
  {author} {\bibfnamefont {C.~R.}\ \bibnamefont {Conner}}, \bibinfo {author}
  {\bibfnamefont {E.}~\bibnamefont {Dumur}}, \bibinfo {author} {\bibfnamefont
  {J.}~\bibnamefont {Grebel}}, \bibinfo {author} {\bibfnamefont {G.~A.}\
  \bibnamefont {Peairs}}, \bibinfo {author} {\bibfnamefont {R.~G.}\
  \bibnamefont {Povey}},\ and\ \bibinfo {author} {\bibfnamefont {A.~N.}\
  \bibnamefont {Cleland}},\ }\bibfield  {title} {\bibinfo {title}
  {Phonon-mediated quantum state transfer and remote qubit entanglement},\
  }\href {https://doi.org/10.1126/science.aaw8415} {\bibfield  {journal}
  {\bibinfo  {journal} {Science}\ }\textbf {\bibinfo {volume} {364}},\ \bibinfo
  {pages} {368} (\bibinfo {year} {2019})}\BibitemShut {NoStop}%
\bibitem [{\citenamefont {Golter}\ \emph {et~al.}(2016)\citenamefont {Golter},
  \citenamefont {Oo}, \citenamefont {Amezcua}, \citenamefont {Lekavicius},
  \citenamefont {Stewart},\ and\ \citenamefont {Wang}}]{Golter2016}%
  \BibitemOpen
  \bibfield  {author} {\bibinfo {author} {\bibfnamefont {D.~A.}\ \bibnamefont
  {Golter}}, \bibinfo {author} {\bibfnamefont {T.}~\bibnamefont {Oo}}, \bibinfo
  {author} {\bibfnamefont {M.}~\bibnamefont {Amezcua}}, \bibinfo {author}
  {\bibfnamefont {I.}~\bibnamefont {Lekavicius}}, \bibinfo {author}
  {\bibfnamefont {K.~A.}\ \bibnamefont {Stewart}},\ and\ \bibinfo {author}
  {\bibfnamefont {H.}~\bibnamefont {Wang}},\ }\bibfield  {title} {\bibinfo
  {title} {Coupling a surface acoustic wave to an electron spin in diamond via
  a dark state},\ }\href {https://doi.org/10.1103/PhysRevX.6.041060} {\bibfield
   {journal} {\bibinfo  {journal} {Phys.\ Rev.\ X}\ }\textbf {\bibinfo {volume}
  {6}},\ \bibinfo {pages} {041060} (\bibinfo {year} {2016})}\BibitemShut
  {NoStop}%
\bibitem [{\citenamefont {Hermelin}\ \emph {et~al.}(2011)\citenamefont
  {Hermelin}, \citenamefont {Takada}, \citenamefont {Yamamoto}, \citenamefont
  {Tarucha}, \citenamefont {Wieck}, \citenamefont {Saminadayar}, \citenamefont
  {B\"auerle},\ and\ \citenamefont {Meunier}}]{Hermelin2011}%
  \BibitemOpen
  \bibfield  {author} {\bibinfo {author} {\bibfnamefont {S.}~\bibnamefont
  {Hermelin}}, \bibinfo {author} {\bibfnamefont {S.}~\bibnamefont {Takada}},
  \bibinfo {author} {\bibfnamefont {M.}~\bibnamefont {Yamamoto}}, \bibinfo
  {author} {\bibfnamefont {S.}~\bibnamefont {Tarucha}}, \bibinfo {author}
  {\bibfnamefont {A.~D.}\ \bibnamefont {Wieck}}, \bibinfo {author}
  {\bibfnamefont {L.}~\bibnamefont {Saminadayar}}, \bibinfo {author}
  {\bibfnamefont {C.}~\bibnamefont {B\"auerle}},\ and\ \bibinfo {author}
  {\bibfnamefont {T.}~\bibnamefont {Meunier}},\ }\bibfield  {title} {\bibinfo
  {title} {Electrons surfing on a sound wave as a platform for quantum optics
  with flying electrons},\ }\href {https://doi.org/10.1038/nature10416}
  {\bibfield  {journal} {\bibinfo  {journal} {Nature}\ }\textbf {\bibinfo
  {volume} {477}},\ \bibinfo {pages} {435} (\bibinfo {year}
  {2011})}\BibitemShut {NoStop}%
\bibitem [{\citenamefont {McNeil}\ \emph {et~al.}(2011)\citenamefont {McNeil},
  \citenamefont {Kataoka}, \citenamefont {Ford}, \citenamefont {Barnes},
  \citenamefont {Anderson}, \citenamefont {Jones}, \citenamefont {Farrer},\
  and\ \citenamefont {Ritchie}}]{McNeil2011}%
  \BibitemOpen
  \bibfield  {author} {\bibinfo {author} {\bibfnamefont {R.~P.~G.}\
  \bibnamefont {McNeil}}, \bibinfo {author} {\bibfnamefont {M.}~\bibnamefont
  {Kataoka}}, \bibinfo {author} {\bibfnamefont {C.~J.~B.}\ \bibnamefont
  {Ford}}, \bibinfo {author} {\bibfnamefont {C.~H.~W.}\ \bibnamefont {Barnes}},
  \bibinfo {author} {\bibfnamefont {D.}~\bibnamefont {Anderson}}, \bibinfo
  {author} {\bibfnamefont {G.~A.~C.}\ \bibnamefont {Jones}}, \bibinfo {author}
  {\bibfnamefont {I.}~\bibnamefont {Farrer}},\ and\ \bibinfo {author}
  {\bibfnamefont {D.~A.}\ \bibnamefont {Ritchie}},\ }\bibfield  {title}
  {\bibinfo {title} {On-demand single-electron transfer between distant quantum
  dots},\ }\href {https://doi.org/10.1038/nature10444} {\bibfield  {journal}
  {\bibinfo  {journal} {Nature}\ }\textbf {\bibinfo {volume} {477}},\ \bibinfo
  {pages} {439} (\bibinfo {year} {2011})}\BibitemShut {NoStop}%
\bibitem [{\citenamefont {Kuzyk}\ and\ \citenamefont {Wang}(2018)}]{Kuzyk2018}%
  \BibitemOpen
  \bibfield  {author} {\bibinfo {author} {\bibfnamefont {M.~C.}\ \bibnamefont
  {Kuzyk}}\ and\ \bibinfo {author} {\bibfnamefont {H.}~\bibnamefont {Wang}},\
  }\bibfield  {title} {\bibinfo {title} {Scaling phononic quantum networks of
  solid-state spins with closed mechanical subsystems},\ }\href
  {https://doi.org/10.1103/PhysRevX.8.041027} {\bibfield  {journal} {\bibinfo
  {journal} {Phys.\ Rev.\ X}\ }\textbf {\bibinfo {volume} {8}},\ \bibinfo
  {pages} {041027} (\bibinfo {year} {2018})}\BibitemShut {NoStop}%
\bibitem [{\citenamefont {Chu}\ and\ \citenamefont
  {Gr\"oblacher}(2020)}]{Chu2020}%
  \BibitemOpen
  \bibfield  {author} {\bibinfo {author} {\bibfnamefont {Y.}~\bibnamefont
  {Chu}}\ and\ \bibinfo {author} {\bibfnamefont {S.}~\bibnamefont
  {Gr\"oblacher}},\ }\bibfield  {title} {\bibinfo {title} {A perspective on
  hybrid quantum opto- and electromechanical systems},\ }\href
  {https://doi.org/10.1063/5.0021088} {\bibfield  {journal} {\bibinfo
  {journal} {Appl.\ Phys.\ Lett.}\ }\textbf {\bibinfo {volume} {117}},\
  \bibinfo {pages} {150503} (\bibinfo {year} {2020})}\BibitemShut {NoStop}%
\bibitem [{\citenamefont {Gustafsson}\ \emph {et~al.}(2012)\citenamefont
  {Gustafsson}, \citenamefont {Santos}, \citenamefont {Johansson},\ and\
  \citenamefont {Delsing}}]{Gustafsson2012}%
  \BibitemOpen
  \bibfield  {author} {\bibinfo {author} {\bibfnamefont {M.~V.}\ \bibnamefont
  {Gustafsson}}, \bibinfo {author} {\bibfnamefont {P.~V.}\ \bibnamefont
  {Santos}}, \bibinfo {author} {\bibfnamefont {G.}~\bibnamefont {Johansson}},\
  and\ \bibinfo {author} {\bibfnamefont {P.}~\bibnamefont {Delsing}},\
  }\bibfield  {title} {\bibinfo {title} {Local probing of propagating acoustic
  waves in a gigahertz echo chamber},\ }\href
  {https://doi.org/10.1038/nphys2217} {\bibfield  {journal} {\bibinfo
  {journal} {Nature Phys.}\ }\textbf {\bibinfo {volume} {8}},\ \bibinfo {pages}
  {338} (\bibinfo {year} {2012})}\BibitemShut {NoStop}%
\bibitem [{\citenamefont {Gustafsson}\ \emph {et~al.}(2014)\citenamefont
  {Gustafsson}, \citenamefont {Aref}, \citenamefont {Kockum}, \citenamefont
  {Ekstr{\"o}m}, \citenamefont {Johansson},\ and\ \citenamefont
  {Delsing}}]{Gustafsson2014}%
  \BibitemOpen
  \bibfield  {author} {\bibinfo {author} {\bibfnamefont {M.~V.}\ \bibnamefont
  {Gustafsson}}, \bibinfo {author} {\bibfnamefont {T.}~\bibnamefont {Aref}},
  \bibinfo {author} {\bibfnamefont {A.~F.}\ \bibnamefont {Kockum}}, \bibinfo
  {author} {\bibfnamefont {M.~K.}\ \bibnamefont {Ekstr{\"o}m}}, \bibinfo
  {author} {\bibfnamefont {G.}~\bibnamefont {Johansson}},\ and\ \bibinfo
  {author} {\bibfnamefont {P.}~\bibnamefont {Delsing}},\ }\bibfield  {title}
  {\bibinfo {title} {Propagating phonons coupled to an artificial atom},\
  }\href {https://doi.org/10.1126/science.1257219} {\bibfield  {journal}
  {\bibinfo  {journal} {Science}\ }\textbf {\bibinfo {volume} {346}},\ \bibinfo
  {pages} {207} (\bibinfo {year} {2014})}\BibitemShut {NoStop}%
\bibitem [{\citenamefont {Patel}\ \emph {et~al.}(2018)\citenamefont {Patel},
  \citenamefont {Wang}, \citenamefont {Jiang}, \citenamefont {Sarabalis},
  \citenamefont {Hill},\ and\ \citenamefont {Safavi-Naeini}}]{Patel2018}%
  \BibitemOpen
  \bibfield  {author} {\bibinfo {author} {\bibfnamefont {R.~N.}\ \bibnamefont
  {Patel}}, \bibinfo {author} {\bibfnamefont {Z.}~\bibnamefont {Wang}},
  \bibinfo {author} {\bibfnamefont {W.}~\bibnamefont {Jiang}}, \bibinfo
  {author} {\bibfnamefont {C.~J.}\ \bibnamefont {Sarabalis}}, \bibinfo {author}
  {\bibfnamefont {J.~T.}\ \bibnamefont {Hill}},\ and\ \bibinfo {author}
  {\bibfnamefont {A.~H.}\ \bibnamefont {Safavi-Naeini}},\ }\bibfield  {title}
  {\bibinfo {title} {Single-mode phononic wire},\ }\href
  {https://doi.org/10.1103/PhysRevLett.121.040501} {\bibfield  {journal}
  {\bibinfo  {journal} {Phys.\ Rev.\ Lett.}\ }\textbf {\bibinfo {volume}
  {121}},\ \bibinfo {pages} {040501} (\bibinfo {year} {2018})}\BibitemShut
  {NoStop}%
\bibitem [{\citenamefont {Fang}\ \emph {et~al.}(2016)\citenamefont {Fang},
  \citenamefont {Matheny}, \citenamefont {Luan},\ and\ \citenamefont
  {Painter}}]{Fang2016}%
  \BibitemOpen
  \bibfield  {author} {\bibinfo {author} {\bibfnamefont {K.}~\bibnamefont
  {Fang}}, \bibinfo {author} {\bibfnamefont {M.~H.}\ \bibnamefont {Matheny}},
  \bibinfo {author} {\bibfnamefont {X.}~\bibnamefont {Luan}},\ and\ \bibinfo
  {author} {\bibfnamefont {O.}~\bibnamefont {Painter}},\ }\bibfield  {title}
  {\bibinfo {title} {Optical transduction and routing of microwave phonons in
  cavity-optomechanical circuits},\ }\href
  {https://doi.org/10.1038/nphoton.2016.107} {\bibfield  {journal} {\bibinfo
  {journal} {Nature Photon.}\ }\textbf {\bibinfo {volume} {10}},\ \bibinfo
  {pages} {489} (\bibinfo {year} {2016})}\BibitemShut {NoStop}%
\bibitem [{\citenamefont {Farrera}\ \emph {et~al.}(2018)\citenamefont
  {Farrera}, \citenamefont {Heinze},\ and\ \citenamefont {{De
  Riedmatten}}}]{Farrera2018}%
  \BibitemOpen
  \bibfield  {author} {\bibinfo {author} {\bibfnamefont {P.}~\bibnamefont
  {Farrera}}, \bibinfo {author} {\bibfnamefont {G.}~\bibnamefont {Heinze}},\
  and\ \bibinfo {author} {\bibfnamefont {H.}~\bibnamefont {{De Riedmatten}}},\
  }\bibfield  {title} {\bibinfo {title} {{Entanglement between a Photonic
  Time-Bin Qubit and a Collective Atomic Spin Excitation}},\ }\href
  {https://doi.org/10.1103/PhysRevLett.120.100501} {\bibfield  {journal}
  {\bibinfo  {journal} {Phys.\ Rev.\ Lett.}\ }\textbf {\bibinfo {volume}
  {120}},\ \bibinfo {pages} {100501} (\bibinfo {year} {2018})}\BibitemShut
  {NoStop}%
\bibitem [{\citenamefont {Chan}\ \emph {et~al.}(2012)\citenamefont {Chan},
  \citenamefont {Safavi-Naeini}, \citenamefont {Hill}, \citenamefont
  {Meenehan},\ and\ \citenamefont {Painter}}]{Chan2012}%
  \BibitemOpen
  \bibfield  {author} {\bibinfo {author} {\bibfnamefont {J.}~\bibnamefont
  {Chan}}, \bibinfo {author} {\bibfnamefont {A.~H.}\ \bibnamefont
  {Safavi-Naeini}}, \bibinfo {author} {\bibfnamefont {J.~T.}\ \bibnamefont
  {Hill}}, \bibinfo {author} {\bibfnamefont {S.}~\bibnamefont {Meenehan}},\
  and\ \bibinfo {author} {\bibfnamefont {O.}~\bibnamefont {Painter}},\
  }\bibfield  {title} {\bibinfo {title} {Optimized optomechanical crystal
  cavity with acoustic radiation shield},\ }\href
  {https://doi.org/10.1063/1.4747726} {\bibfield  {journal} {\bibinfo
  {journal} {App. Phys. Lett.}\ }\textbf {\bibinfo {volume} {101}},\ \bibinfo
  {pages} {081115} (\bibinfo {year} {2012})}\BibitemShut {NoStop}%
\bibitem [{\citenamefont {Hong}\ \emph {et~al.}(2017)\citenamefont {Hong},
  \citenamefont {Riedinger}, \citenamefont {Marinkovi\'{c}}, \citenamefont
  {Wallucks}, \citenamefont {Hofer}, \citenamefont {Norte}, \citenamefont
  {Aspelmeyer},\ and\ \citenamefont {Gr\"oblacher}}]{Hong2017}%
  \BibitemOpen
  \bibfield  {author} {\bibinfo {author} {\bibfnamefont {S.}~\bibnamefont
  {Hong}}, \bibinfo {author} {\bibfnamefont {R.}~\bibnamefont {Riedinger}},
  \bibinfo {author} {\bibfnamefont {I.}~\bibnamefont {Marinkovi\'{c}}},
  \bibinfo {author} {\bibfnamefont {A.}~\bibnamefont {Wallucks}}, \bibinfo
  {author} {\bibfnamefont {S.~G.}\ \bibnamefont {Hofer}}, \bibinfo {author}
  {\bibfnamefont {R.~A.}\ \bibnamefont {Norte}}, \bibinfo {author}
  {\bibfnamefont {M.}~\bibnamefont {Aspelmeyer}},\ and\ \bibinfo {author}
  {\bibfnamefont {S.}~\bibnamefont {Gr\"oblacher}},\ }\bibfield  {title}
  {\bibinfo {title} {Hanbury {B}rown and {T}wiss interferometry of single
  phonons from an optomechanical resonator},\ }\href
  {https://doi.org/10.1126/science.aan7939} {\bibfield  {journal} {\bibinfo
  {journal} {Science}\ }\textbf {\bibinfo {volume} {358}},\ \bibinfo {pages}
  {203} (\bibinfo {year} {2017})}\BibitemShut {NoStop}%
\bibitem [{\citenamefont {Qiu}\ \emph {et~al.}(2020)\citenamefont {Qiu},
  \citenamefont {Shomroni}, \citenamefont {Seidler},\ and\ \citenamefont
  {Kippenberg}}]{Qiu2020}%
  \BibitemOpen
  \bibfield  {author} {\bibinfo {author} {\bibfnamefont {L.}~\bibnamefont
  {Qiu}}, \bibinfo {author} {\bibfnamefont {I.}~\bibnamefont {Shomroni}},
  \bibinfo {author} {\bibfnamefont {P.}~\bibnamefont {Seidler}},\ and\ \bibinfo
  {author} {\bibfnamefont {T.~J.}\ \bibnamefont {Kippenberg}},\ }\bibfield
  {title} {\bibinfo {title} {Laser cooling of a nanomechanical oscillator to
  its zero-point energy},\ }\href
  {https://doi.org/10.1103/PhysRevLett.124.173601} {\bibfield  {journal}
  {\bibinfo  {journal} {Phys.\ Rev.\ Lett.}\ }\textbf {\bibinfo {volume}
  {124}},\ \bibinfo {pages} {173601} (\bibinfo {year} {2020})}\BibitemShut
  {NoStop}%
\bibitem [{\citenamefont {Riedinger}\ \emph {et~al.}(2016)\citenamefont
  {Riedinger}, \citenamefont {Hong}, \citenamefont {Norte}, \citenamefont
  {Slater}, \citenamefont {Shang}, \citenamefont {Krause}, \citenamefont
  {Anant}, \citenamefont {Aspelmeyer},\ and\ \citenamefont
  {Gr\"oblacher}}]{Riedinger2016}%
  \BibitemOpen
  \bibfield  {author} {\bibinfo {author} {\bibfnamefont {R.}~\bibnamefont
  {Riedinger}}, \bibinfo {author} {\bibfnamefont {S.}~\bibnamefont {Hong}},
  \bibinfo {author} {\bibfnamefont {R.~A.}\ \bibnamefont {Norte}}, \bibinfo
  {author} {\bibfnamefont {J.~A.}\ \bibnamefont {Slater}}, \bibinfo {author}
  {\bibfnamefont {J.}~\bibnamefont {Shang}}, \bibinfo {author} {\bibfnamefont
  {A.~G.}\ \bibnamefont {Krause}}, \bibinfo {author} {\bibfnamefont
  {V.}~\bibnamefont {Anant}}, \bibinfo {author} {\bibfnamefont
  {M.}~\bibnamefont {Aspelmeyer}},\ and\ \bibinfo {author} {\bibfnamefont
  {S.}~\bibnamefont {Gr\"oblacher}},\ }\bibfield  {title} {\bibinfo {title}
  {Non-classical correlations between single photons and phonons from a
  mechanical oscillator},\ }\href {https://doi.org/10.1038/nature16536}
  {\bibfield  {journal} {\bibinfo  {journal} {Nature}\ }\textbf {\bibinfo
  {volume} {530}},\ \bibinfo {pages} {313} (\bibinfo {year}
  {2016})}\BibitemShut {NoStop}%
\bibitem [{\citenamefont {Weis}\ \emph {et~al.}(2010)\citenamefont {Weis},
  \citenamefont {Rivi\`ere}, \citenamefont {Del\'eglise}, \citenamefont
  {Gavartin}, \citenamefont {Arcizet}, \citenamefont {Schliesser},\ and\
  \citenamefont {Kippenberg}}]{Weis2010}%
  \BibitemOpen
  \bibfield  {author} {\bibinfo {author} {\bibfnamefont {S.}~\bibnamefont
  {Weis}}, \bibinfo {author} {\bibfnamefont {R.}~\bibnamefont {Rivi\`ere}},
  \bibinfo {author} {\bibfnamefont {S.}~\bibnamefont {Del\'eglise}}, \bibinfo
  {author} {\bibfnamefont {E.}~\bibnamefont {Gavartin}}, \bibinfo {author}
  {\bibfnamefont {O.}~\bibnamefont {Arcizet}}, \bibinfo {author} {\bibfnamefont
  {A.}~\bibnamefont {Schliesser}},\ and\ \bibinfo {author} {\bibfnamefont
  {T.~J.}\ \bibnamefont {Kippenberg}},\ }\bibfield  {title} {\bibinfo {title}
  {{Optomechanically Induced Transparency}},\ }\href
  {https://doi.org/10.1126/science.1195596} {\bibfield  {journal} {\bibinfo
  {journal} {Science}\ }\textbf {\bibinfo {volume} {330}},\ \bibinfo {pages}
  {1520} (\bibinfo {year} {2010})}\BibitemShut {NoStop}%
\bibitem [{\citenamefont {Meenehan}\ \emph {et~al.}(2014)\citenamefont
  {Meenehan}, \citenamefont {Cohen}, \citenamefont {Gr{\"{o}}blacher},
  \citenamefont {Hill}, \citenamefont {Safavi-Naeini}, \citenamefont
  {Aspelmeyer},\ and\ \citenamefont {Painter}}]{Meenehan2014}%
  \BibitemOpen
  \bibfield  {author} {\bibinfo {author} {\bibfnamefont {S.~M.}\ \bibnamefont
  {Meenehan}}, \bibinfo {author} {\bibfnamefont {J.~D.}\ \bibnamefont {Cohen}},
  \bibinfo {author} {\bibfnamefont {S.}~\bibnamefont {Gr{\"{o}}blacher}},
  \bibinfo {author} {\bibfnamefont {J.~T.}\ \bibnamefont {Hill}}, \bibinfo
  {author} {\bibfnamefont {A.~H.}\ \bibnamefont {Safavi-Naeini}}, \bibinfo
  {author} {\bibfnamefont {M.}~\bibnamefont {Aspelmeyer}},\ and\ \bibinfo
  {author} {\bibfnamefont {O.}~\bibnamefont {Painter}},\ }\bibfield  {title}
  {\bibinfo {title} {{Silicon optomechanical crystal resonator at millikelvin
  temperatures}},\ }\href {https://doi.org/10.1103/PhysRevA.90.011803}
  {\bibfield  {journal} {\bibinfo  {journal} {Phys.\ Rev.\ A}\ }\textbf
  {\bibinfo {volume} {90}},\ \bibinfo {pages} {011803} (\bibinfo {year}
  {2014})}\BibitemShut {NoStop}%
\bibitem [{\citenamefont {Pang}\ \emph {et~al.}(2020)\citenamefont {Pang},
  \citenamefont {Yang}, \citenamefont {Dou}, \citenamefont {Li}, \citenamefont
  {Zhang}, \citenamefont {Poem}, \citenamefont {Saunders}, \citenamefont
  {Tang}, \citenamefont {Nunn}, \citenamefont {Walmsley},\ and\ \citenamefont
  {Jin}}]{Pang2020}%
  \BibitemOpen
  \bibfield  {author} {\bibinfo {author} {\bibfnamefont {X.-L.}\ \bibnamefont
  {Pang}}, \bibinfo {author} {\bibfnamefont {A.-L.}\ \bibnamefont {Yang}},
  \bibinfo {author} {\bibfnamefont {J.-P.}\ \bibnamefont {Dou}}, \bibinfo
  {author} {\bibfnamefont {H.}~\bibnamefont {Li}}, \bibinfo {author}
  {\bibfnamefont {C.-N.}\ \bibnamefont {Zhang}}, \bibinfo {author}
  {\bibfnamefont {E.}~\bibnamefont {Poem}}, \bibinfo {author} {\bibfnamefont
  {D.~J.}\ \bibnamefont {Saunders}}, \bibinfo {author} {\bibfnamefont
  {H.}~\bibnamefont {Tang}}, \bibinfo {author} {\bibfnamefont {J.}~\bibnamefont
  {Nunn}}, \bibinfo {author} {\bibfnamefont {I.~A.}\ \bibnamefont {Walmsley}},\
  and\ \bibinfo {author} {\bibfnamefont {X.-M.}\ \bibnamefont {Jin}},\
  }\bibfield  {title} {\bibinfo {title} {A hybrid quantum memory-enabled
  network at room temperature},\ }\href
  {https://doi.org/10.1126/sciadv.aax1425} {\bibfield  {journal} {\bibinfo
  {journal} {Sci.\ Adv.}\ }\textbf {\bibinfo {volume} {6}},\ \bibinfo {pages}
  {eaax1425} (\bibinfo {year} {2020})}\BibitemShut {NoStop}%
\bibitem [{\citenamefont {Borselli}\ \emph {et~al.}(2006)\citenamefont
  {Borselli}, \citenamefont {Johnson},\ and\ \citenamefont
  {Painter}}]{Borselli2006}%
  \BibitemOpen
  \bibfield  {author} {\bibinfo {author} {\bibfnamefont {M.}~\bibnamefont
  {Borselli}}, \bibinfo {author} {\bibfnamefont {T.~J.}\ \bibnamefont
  {Johnson}},\ and\ \bibinfo {author} {\bibfnamefont {O.}~\bibnamefont
  {Painter}},\ }\bibfield  {title} {\bibinfo {title} {Measuring the role of
  surface chemistry in silicon microphotonics},\ }\href
  {https://doi.org/10.1063/1.2191475} {\bibfield  {journal} {\bibinfo
  {journal} {Appl.\ Phys.\ Lett.}\ }\textbf {\bibinfo {volume} {88}},\ \bibinfo
  {pages} {131114} (\bibinfo {year} {2006})}\BibitemShut {NoStop}%
\bibitem [{\citenamefont {Gr\"oblacher}\ \emph {et~al.}(2013)\citenamefont
  {Gr\"oblacher}, \citenamefont {Hill}, \citenamefont {Safavi-Naeini},
  \citenamefont {Chan},\ and\ \citenamefont {Painter}}]{Groeblacher2013a}%
  \BibitemOpen
  \bibfield  {author} {\bibinfo {author} {\bibfnamefont {S.}~\bibnamefont
  {Gr\"oblacher}}, \bibinfo {author} {\bibfnamefont {J.~T.}\ \bibnamefont
  {Hill}}, \bibinfo {author} {\bibfnamefont {A.~H.}\ \bibnamefont
  {Safavi-Naeini}}, \bibinfo {author} {\bibfnamefont {J.}~\bibnamefont
  {Chan}},\ and\ \bibinfo {author} {\bibfnamefont {O.}~\bibnamefont
  {Painter}},\ }\bibfield  {title} {\bibinfo {title} {Highly efficient coupling
  from an optical fiber to a nanoscale silicon optomechanical cavity},\ }\href
  {https://doi.org/10.1063/1.4826924} {\bibfield  {journal} {\bibinfo
  {journal} {Appl.\ Phys.\ Lett.,}\ }\textbf {\bibinfo {volume} {103}},\
  \bibinfo {pages} {181104} (\bibinfo {year} {2013})}\BibitemShut {NoStop}%
\bibitem [{\citenamefont {Lebreton}\ \emph {et~al.}(2013)\citenamefont
  {Lebreton}, \citenamefont {Abram}, \citenamefont {Takemura}, \citenamefont
  {Kuwata-Gonokami}, \citenamefont {Robert-Philip},\ and\ \citenamefont
  {Beveratos}}]{Lebreton2013}%
  \BibitemOpen
  \bibfield  {author} {\bibinfo {author} {\bibfnamefont {A.}~\bibnamefont
  {Lebreton}}, \bibinfo {author} {\bibfnamefont {I.}~\bibnamefont {Abram}},
  \bibinfo {author} {\bibfnamefont {N.}~\bibnamefont {Takemura}}, \bibinfo
  {author} {\bibfnamefont {M.}~\bibnamefont {Kuwata-Gonokami}}, \bibinfo
  {author} {\bibfnamefont {I.}~\bibnamefont {Robert-Philip}},\ and\ \bibinfo
  {author} {\bibfnamefont {A.}~\bibnamefont {Beveratos}},\ }\bibfield  {title}
  {\bibinfo {title} {Stochastically sustained population oscillations in
  high-$\beta$ nanolasers},\ }\href
  {https://doi.org/10.1088/1367-2630/15/3/033039} {\bibfield  {journal}
  {\bibinfo  {journal} {New J.\ Phys.}\ }\textbf {\bibinfo {volume} {15}},\
  \bibinfo {pages} {033039} (\bibinfo {year} {2013})}\BibitemShut {NoStop}%
\end{thebibliography}
\end{document}